\documentclass{imamat}

\usepackage{graphicx}

\newcommand{\mylab}[1]{\label{#1}}
\renewcommand{\vec}{\mathbf}

\jno{iarxxx}
\received{XX XX XXXX}
\revised{XX XX XXXX}
\accepted{XX XX XXXX}

\begin{document}

\title{Localized states in passive and active phase-field-crystal models}

\shorttitle{Localized states in passive and active phase-field-crystal models} 
\shortauthorlist{M. P. Holl \textit{et al.}} 

\author{
\name{Max Philipp Holl\thanks{ORCID ID: 0000-0001-6451-9723}}
\address{Institut f\"ur Theoretische Physik, Westf\"alische Wilhelms-Universit\"at M\"unster, Wilhelm-Klemm-Str.\ 9, 48149 M\"unster, Germany\email{Corresponding author: m.p.holl@wwu.de}}
\name{Andrew J. Archer\thanks{ORCID ID: 0000-0002-4706-2204}}
\address{Department of Mathematical Sciences, Loughborough University, Loughborough, Leicestershire, LE11 3TU, UK\\Interdisciplinary Centre for Mathematical Modelling, Loughborough University, Loughborough, Leicestershire, LE11 3TU, UK}
\name{Svetlana V. Gurevich\thanks{ORCID: 0000-0002-5101-4686}}
\address{Institut f\"ur Theoretische Physik, Westf\"alische Wilhelms-Universit\"at M\"unster, Wilhelm-Klemm-Str.\ 9, 48149 M\"unster, Germany\\ Center for Nonlinear Science (CeNoS), Westf{\"a}lische Wilhelms-Universit\"at M\"unster, Corrensstr.\ 2, 48149 M\"unster, Germany}
\name{Edgar Knobloch}
\address{Department of Physics, University of California, Berkeley, California 94720, USA}
\name{Lukas Ophaus}
\address{Institut f\"ur Theoretische Physik, Westf\"alische Wilhelms-Universit\"at M\"unster, Wilhelm-Klemm-Str.\ 9, 48149 M\"unster, Germany}
\and
\name{Uwe Thiele\thanks{homepage: http://www.uwethiele.de; ORCID ID: 0000-0001-7989-9271}}
\address{Institut f\"ur Theoretische Physik, Westf\"alische Wilhelms-Universit\"at M\"unster, Wilhelm-Klemm-Str.\ 9, 48149 M\"unster, Germany~\\
  Center for Nonlinear Science (CeNoS), Westf{\"a}lische Wilhelms-Universit\"at M\"unster, Corrensstr.\ 2, 48149 M\"unster, Germany
  \email{u.thiele@uni-muenster.de}}
}
\maketitle

\begin{abstract}
{The passive conserved Swift-Hohenberg equation (or phase-field-crystal [PFC] model) corresponds to a gradient dynamics for a single order parameter field related to density. It provides a simple microscopic description of the thermodynamic transition between liquid and crystalline states. In addition to spatially extended periodic structures, the model describes a large variety of steady spatially localized structures. In appropriate bifurcation diagrams the corresponding solution branches exhibit characteristic slanted homoclinic snaking. In an active PFC model, encoding for instance the active motion of self-propelled colloidal particles, the gradient dynamics structure is broken by a coupling between density and an additional polarization field. Then, resting and traveling localized states are found with transitions characterized by parity-breaking drift bifurcations. Here, we first briefly review the snaking behavior of localized states in passive and active PFC models before discussing the bifurcation behavior of localized states in systems of (i) two coupled passive PFC equations described by common gradient dynamics, (ii) two coupled passive PFC where the coupling breaks the gradient dynamics structure, and (iii) a passive PFC coupled to an active PFC.}
{Localized States, Homoclinic Slanted Snaking, active Phase Field Crystal, Bifurcation Analysis, Numerical Continuation}
\end{abstract}

\section{Introduction} \mylab{sec:intro}
%
Phenomena involving pattern formation are widespread in nature and are also encountered in technologically relevant processes \citep{Ball1999}. Strictly speaking, by ``pattern formation'' one refers to the self-organized emergence\footnote{%
	Here, we use ``self-organization'' in a loose sense, but acknowledge that its precise meaning, exact relation to dissipative processes and discrimination from ``(dynamic) self-assembly'' are widely and wildly discussed issues.} of spatially periodic (or quasiperiodic) structures that can be steady or, more generally, exhibit temporally regular dynamics \citep{CrHo1993rmp,Hoyle2006,Pismen2006,CrossGreenside2009}. Often, the underlying processes are described by continuum models, i.e., the patterns correspond to solutions of partial differential equations (PDE).
Over the years much attention has focused on infinitely extended patterns and on patterns in finite domains, where the imposed confinement provides particular boundary conditions \citep{CrHo1993rmp}. However, it was also found that under certain conditions finite pattern patches may coexist with an infinitely extended uniform background (or a background formed by a different pattern). Such structures are the localized states that are the subject of the present volume. Examples include the localized buckling of a long beam under compression \citep{HPCW2000nd}, localized patches of traveling waves in binary mixture convection \citep{BaLK1991prl,SOYK1991pra}, dissipative solitons in bistable optical systems \citep{TlML1994prl}, bound states of oscillons in vibrated layers of sand \citep{UmMS1996n}, and patches of peaks close to the onset of the Rosensweig instability of the free-surface of a magnetic fluid \citep{LGRR2015jfm}. 

Localized states have been extensively investigated for many individual physical systems like those mentioned above. Remarkably, these share a number of common features that are independent of the particular physics involved. In bifurcation diagrams that present the various localized states that arise as a function of a control parameter, the branches of localized states of different types often form a snakes-and-ladders structure referred to as ``homoclinic snaking'' \citep{WoCh1999pd,BuKn2006pre}.

Localized states are frequently described by means of generic equations from the theory of pattern formation, particularly the Swift-Hohenberg equation and its variants \citep{HMBD1995pre,SaBr1996pd,BuKn2006pre,BuKn2007pla}. This equation can be written as
\begin{equation}
\partial_t \phi = -M \frac{\delta \mathcal{F}_\mathrm{sh}[\phi]}{\delta \phi},
\mylab{eq:sh}
\end{equation}
where $M$ is a positive mobility constant. This variational form highlights the fact that the Swift-Hohenberg equation corresponds to a gradient dynamics for the order parameter field $\phi$ with the underlying energy functional
\begin{eqnarray}
\mathcal{F}_\mathrm{sh}[\phi]\equiv \int_V \mathrm{d}^n {\bf r} \left[f(\phi) - q^2(\nabla\phi)^2 + \frac{1}{2} (\Delta\phi)^2\right].
\mylab{eq:sh-en}
\end{eqnarray}
Since equation~(\ref{eq:sh}) does not preserve the ``mass'' $\int_V \phi\,\mathrm{d}^n {\bf r}$, it represents a ``nonconserved dynamics''. 

The local energy $f(\phi)$ is a double-well potential, e.g., the simplest form $f=f_{24}\equiv(r+q^2)\phi^2/2+\phi^4/4$ \citep{CrHo1993rmp}, or $f=f_{234}\equiv(r+q^2)\phi^2/2-b_3\phi^3/3+\phi^4/4$ \citep{HMBD1995pre,BuKn2006pre}.\footnote{Note that different sign conventions for the parameter $r$ are in use.} More complicated forms like $f=f_{246}\equiv(r+q^2)\phi^2/2-b_4\phi^4/4+\phi^6/6$ \citep{SaBr1996pd,BuKn2007pla,ALBK2010sjads,HoKn2011pre} and a quadratic-quartic-sextic-octic potential \citep{KnUW2019pre} have also been investigated. Snaking localized states are found in all of the examples mentioned except $f_{24}$ and have been investigated in one \citep{HoKn2011pre}, two \citep{LSAC2008sjads,LlSa2009n,ALBK2010sjads,McSa2010pd} and three dimensions \citep{McSa2010pd}. The influence of various nonvariational modifications of Eq.~(\ref{eq:sh}) through additional terms that break its gradient structure has also been discussed \citep{KuHP1996pre,BuHK2009pre,HoKn2011pre,KoTl2007c,BuDa2012sjads,LeTL2004ijqc,BCCL2016sr,HBCP2019pra}.

The simple potential $f_{24}$ does not allow for localized states because the primary bifurcation is always supercritical and, in consequence, the trivial uniform state $\phi=0$ and the spatially periodic patterned state do not exist as linearly stable states at identical parameter values.  Such coexistence is a condition for localized states to occur in systems with nonconserved dynamics. 
The resulting localized states exhibit aligned or ``vertical'' snaking where like saddle-node bifurcations all occur asymptotically at the same parameter value \citep{BuKn2006pre,BuKn2007pla}.

In contrast to the Swift-Hohenberg equation (\ref{eq:sh}), the conserved Swift-Hohenberg equation \citep{TARG2013pre} does conserve mass. This equation is prominent in material science as the ``phase-field-crystal (PFC) model'' \citep{ElGr2004pre,ELWG2012ap} and is sometimes called the derivative Swift-Hohenberg equation \citep{MaCo2000n,Cox2004pla}. It takes the form
\begin{equation}
\partial_t \phi = M \Delta\frac{\delta \mathcal{F}_\mathrm{sh}[\phi]}{\delta \phi}
\mylab{eq:csh}
\end{equation}
with the energy functional again given by Eq.~(\ref{eq:sh-en}). In this case the local energy $f(\phi)$ can be taken to be the simplest double-well potential $f_{24}$ since this time it does allow for the coexistence of homogeneous and patterned states even when the primary bifurcation is supercritical: because of the mass conservation, the coexisting states typically represent different mean concentrations \citep{TARG2013pre}. In effect, the overall mean density $\bar\phi$ takes the role of the parameter $b_3$ in $f_{234}$ in the standard Swift-Hohenberg equation.

One finds that the localized states again form a homoclinic snaking structure, although this time it is no longer aligned but slanted \citep{TARG2013pre}. Such slanted snaking is also observed in other pattern-forming systems with a conservation law \citep{Dawe2008sjads,LoBK2011jfm,Knob2015arcmp,PACF2017prf}. Ultimately, the tilting of the snakes-and-ladders structure is due to the finite domain size as expulsion [absorption] of mass from [into] the expanding localized pattern implies an increase [decrease] of the mean density in the uniform background. This progressively changes the value of the control parameter where the next peak of the pattern forms. As a result, localized states exist in a larger parameter range than in nonconserved systems. Slanted snaking may also appear in systems with spatial periodic heterogeneity \citep{BoCR2008pre} and be induced by certain boundary conditions \citep{KoAC2009prl}.

For the conserved Swift-Hohenberg equation and related PFC models, slanted snaking is observed when using the mean density or effective temperature $r$ as the primary control parameter \citep{RATK2012pre,TARG2013pre}. However, it takes the form of standard aligned snaking when the bifurcation diagram is plotted as a function of the chemical potential. Essentially, this results from the mean density being a natural control parameter for the conserved dynamics while controlling the chemical potential corresponds to the nonconserved dynamics (see the conclusion by \citet{TARG2013pre} and~\citet{EGUW2019springer}). The localization of patterns due to a conserved quantity is also considered by \citet{CoMa2003pd} for a nonvariational extension of the PFC model. There, in the nomenclature of \citet{EGUW2019springer}, a nonvariational chemical potential and an additional nonvariational flux are added to the conserved Swift-Hohenberg equation. 

In a major extension of the PFC model, it may be combined with elements of the Toner-Tu theory for self-propelled particles \citep{ToTu1995prl} to obtain an active PFC model as a description of crystallization-like nonequilibrium transitions in systems of active colloidal particles \citep{MeLo2013prl}. In this approach the density field is coupled to a new field, the polarization field. The coupling, quantified by an ``activity parameter'', breaks the gradient dynamics structure of the PFC model thereby allowing for sustained motion and states that oscillate in time. Note that nonvariational modifications of the standard SH equation may also result in traveling states \citep{KoTl2007c,HoKn2011pre,BuDa2012sjads} although with quite different onset behavior.

The active PFC model is one of many models developed to capture collective phenomena, like motility-induced clustering and swarm formation in active hydrodynamic and soft matter systems \citep{MJRL2013rmp}. It shows transitions between a liquid state and an extended patterned (crystalline) state, as well as transitions between resting and traveling patterns \citep{MeLo2013prl,MeOL2014pre,ChGT2016el}. Experimentally, related resting \citep{ThKu2005fml} or traveling \citep{PSSP2013s,TCPY2012prl,PeWL2015prl,GTDY2018nc,BDLR2016rmp} patches of particles with nearly crystalline order \citep{ToTR2005ap} are reported as ``active crystals'', ``flying crystals'' or ``living crystals''. The activity due to self-propulsion induces properties that differ from passive crystalline clusters. For instance, it can amend the critical density and temperature at which crystallization takes place and may induce translational and rotational motion. 

Recently it has been found that the active PFC model not only allows for periodic states but also for many types of resting and traveling localized states \citep{OpGT2018pre}. There, a detailed analysis of the underlying bifurcation structure in one spatial dimension is provided, with a particular emphasis on the onset of motion. Further details of the bifurcation structure and an investigation of the scattering behavior of traveling localized states are presented by \citet{OKGT2020c}. First results on localized states in two dimensions were presented by \citet{OKGT2020pre}.

In the present contribution we focus on the one-dimensional (1D) case and present an overview and comparison of the behavior of localized states in PFC systems. In particular, section~\ref{sec:pfc} reviews and extends results for the classic passive PFC model, while section~\ref{sec:apfc} gives an overview of the active PFC model. Then, in section~\ref{sec:2pfc} we present the first results for localized states in a passive binary PFC system and, in particular, a symmetric mixture of colloidal particles. The subsequent section~\ref{sec:pfcapfc} considers an active binary PFC system modeling a mixture of passive and active colloids. In all cases we discuss the structure of the model, sketch the linear stability results for uniform states and present typical bifurcation diagrams for steady (and traveling) localized states. The paper concludes with section~\ref{sec:conc}.

\section{Slanted snaking in the passive PFC model}
\mylab{sec:pfc}
%
Introducing the energy (\ref{eq:sh-en}) into the mass-conserving dynamics (\ref{eq:csh}) results in the kinetic equation
\begin{equation}
\partial_t\,\psi\,=\, M \nabla^2\left[r\psi + (\nabla^2+q^2)^2\psi + (\psi+\bar\phi)^3\right],
\mylab{eq:csh}
\end{equation}
where we introduced the field $\psi\equiv\phi-\bar\phi$ with the property $\int_V \psi\,\mathrm{d}^n{\bf r}=0$. Thus $\bar\phi$ becomes a parameter of the equation. In the underlying $\mathcal{F}[\phi]$ the quartic term may be replaced by other types of nonlinearity, such as $f_{234}$ \citep{MaCo2000n,Cox2004pla,TGTP2009prl,TBVL2009pre} without substantial change in behavior. 

Here we consider Eq.~(\ref{eq:csh}) in one dimension only, with $M=1$ and $q=1$, i.e.,
\begin{equation}
\partial_t\,\psi\,=\, \partial^2_x\left[r\psi + (\partial^2_{x}+1)^2\psi + (\psi+\bar\phi)^3\right].
\mylab{eq:csh-loc}
\end{equation}
For results in 2D and 3D, see \citet{TARG2013pre,EGUW2019springer} and~\citet{TFEK2019njp}.

This equation is invariant under the reflection  $x\rightarrow -x$, i.e., it is reversible in space.  Steady states ($\partial_t\,\psi\,=\,0$) are solutions of the fourth order ordinary differential equation
\begin{equation}
0 = r\psi + (\partial^2_{x}+1)^2\psi + (\psi+\bar\phi)^3 - \tilde\mu,
\mylab{eq:csh-loc-steady}
\end{equation}
where $\tilde\mu$ is an integration constant that is directly related to the chemical potential $\mu=\tilde\mu+(1+r)\bar\phi$ if $\mathcal{F}[\phi]$ is a free energy functional \citep[cf.][]{TARG2013pre}. 

\begin{figure}
	\centering
	\includegraphics[width=0.8\hsize]{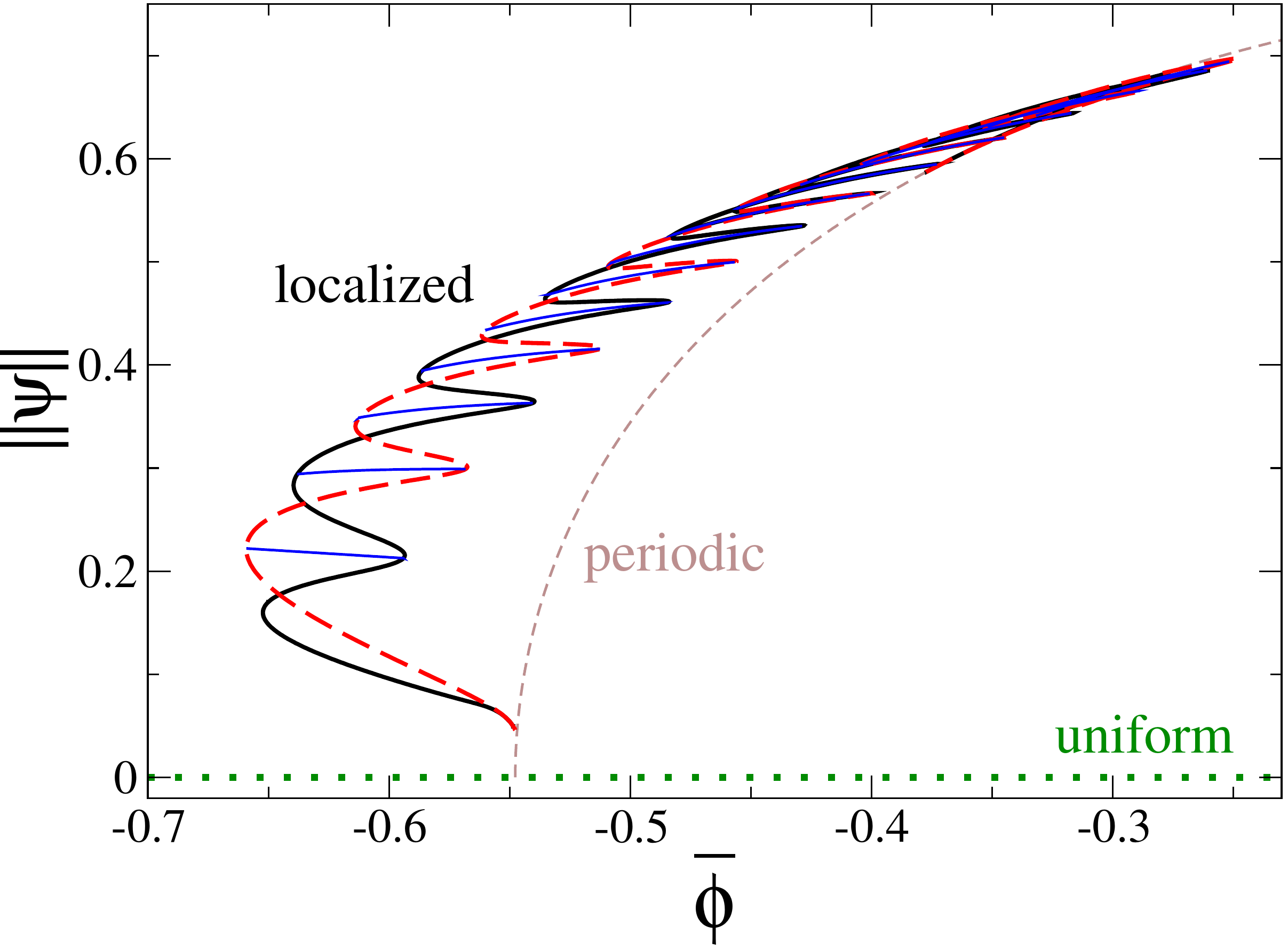}
	\caption{Typical bifurcation diagram for steady states of the conserved Swift-Hohenberg equation (passive PFC model)
		in 1d [Eq.~(\ref{eq:csh-loc})]. It is given in terms of the $L^2$ norm $||\psi||$
		as a function of the mean density $\bar\phi$ for fixed domain size $L=100$ and effective temperature $r=-0.9$.
		The horizontal green dotted line is the trivial uniform (liquid) state $\psi(x)=0$, while the thin brown dashed line corresponds to periodic (crystalline) states. The intertwined solid black and dashed red lines represent slanted homoclinic snaking of symmetric steady localized states with odd (LS$_\mathrm{odd}$) and even (LS$_\mathrm{even}$) number of peaks, respectively. The blue solid lines correspond to asymmetric localized states. Together these form a slanted snakes-and-ladders structure.
	}
	\label{fig:loc-fam-rm09}
\end{figure}

The typical bifurcation behavior as a function of the mean density $\bar\phi $ at a modest effective temperature $r=-0.9$ in the range where crystallization can occur is shown in Fig.~\ref{fig:loc-fam-rm09} using the $L^2$ norm $||\psi||$ as a solution measure. The uniform, i.e., liquid, state exists at all densities. It is linearly stable at large negative $\bar\phi$ and becomes linearly unstable at $\bar\phi\approx-0.55$. A branch of periodic, i.e., crystalline states with $n=16$ peaks for a system of length $L=100$ emerges in a supercritical pitchfork bifurcation and is initially stable. Branches with other peak numbers emerge nearby but are unstable and not relevant here (not shown). The $n=16$ periodic states lose stability almost immediately in a further pitchfork bifurcation where two branches of symmetric localized states bifurcate subcritically. The presence of this bifurcation is a consequence of mass conservation \citep{MaCo2000n}. These localized states are symmetric with respect to reflection in their midpoint where they have a maximum (LS$_\mathrm{odd}$, black solid line) or a minimum (LS$_\mathrm{even}$, dashed red line). Both are unstable until they fold back towards larger $\bar\phi$ in saddle-node bifurcations. The branches then wiggle back and forth exhibiting pronounced slanted snaking until they terminate back on the $n=16$ periodic branch at about $\bar\phi\approx-0.3$ in another subcritical pitchfork bifurcation. Beyond this point the $n=16$ periodic branch is again stable. The two branches of symmetric localized states are connected by 14 branches of steady asymmetric localized states that form the rungs of the snakes-and-ladders structure. For more details, and in particular a discussion of the changes in stability along the branches, see \citet{TARG2013pre}.

\begin{figure}
		\centering
	(a)\includegraphics[width=0.6\hsize]{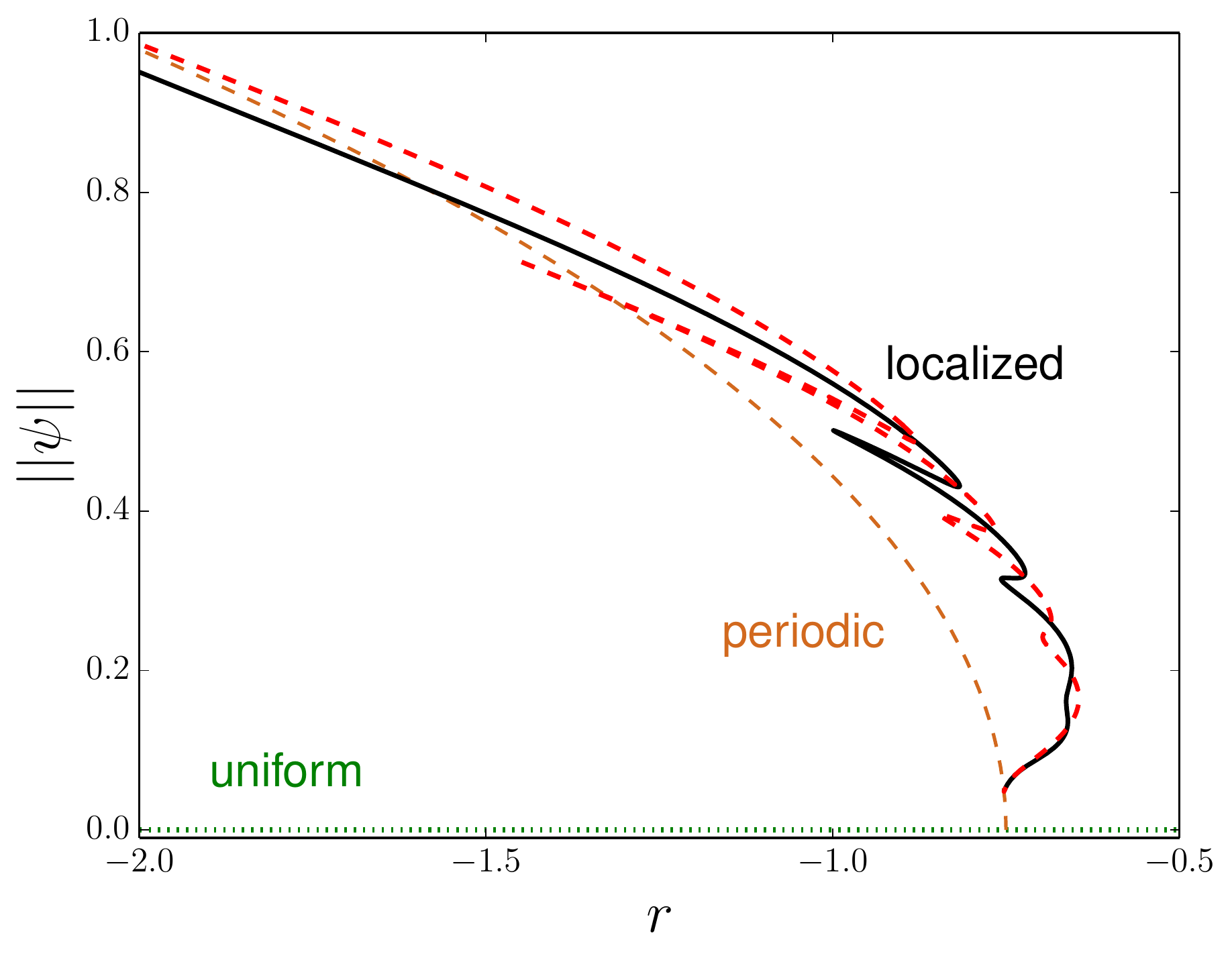}\\
	(b)\includegraphics[width=0.6\hsize]{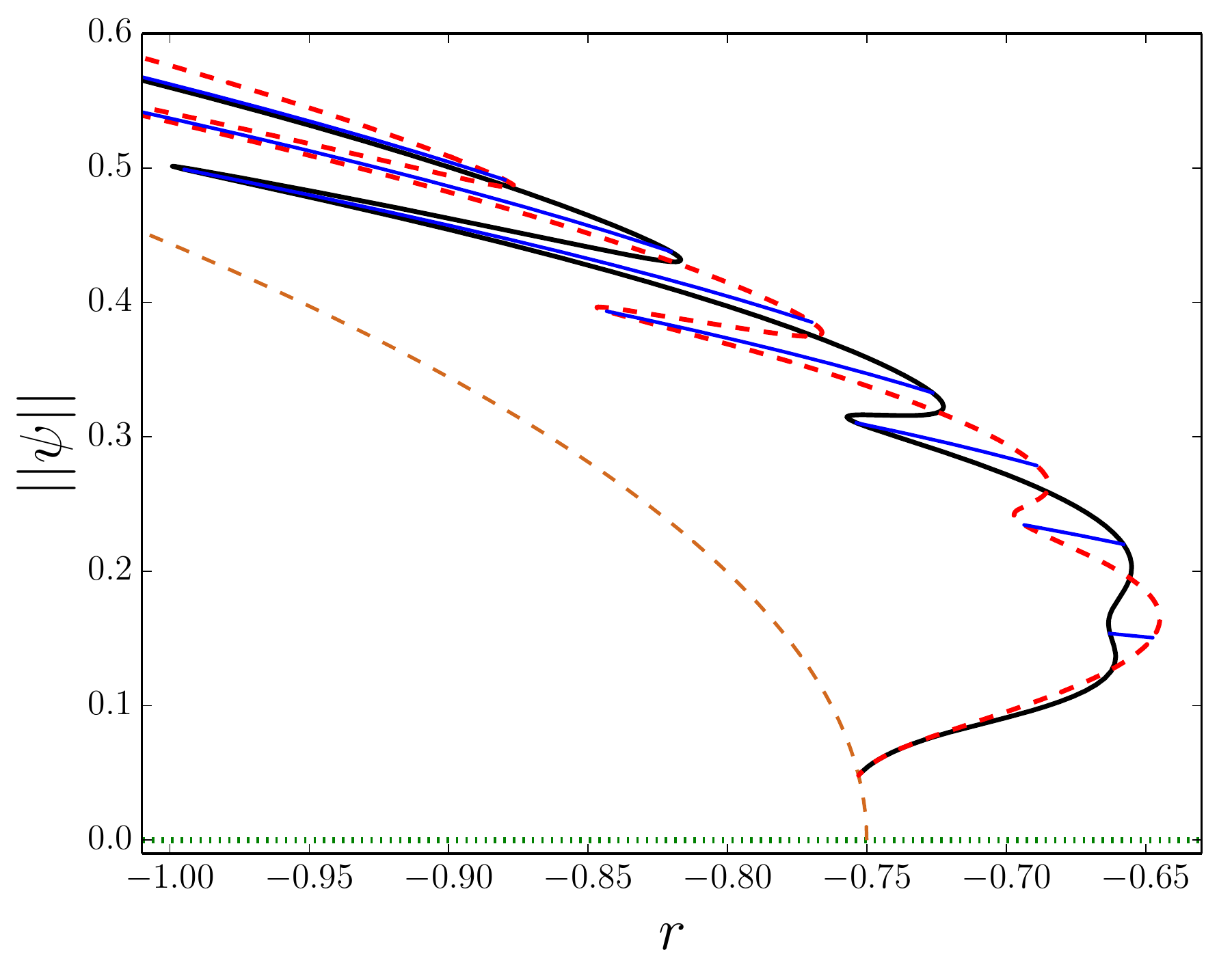}
	\caption{Bifurcation diagram for the passive PFC model [Eq.~(\ref{eq:csh-loc})] using the effective temperature $r$ as the control parameter for fixed $\bar\phi=0.5$. Panel (a) shows the homogeneous, periodic and symmetric localized states. The magnification in panel (b) additionally includes the asymmetric localized states (blue rungs). The remaining parameters and line styles are as in Fig.~\ref{fig:loc-fam-rm09}. }
	\label{fig:loc-fam-phimean05}
\end{figure}

Next, we supplement \citet{TARG2013pre} by discussing the bifurcation behavior when the temperature $r$ is employed as control parameter at fixed mean density. A typical case is presented in Fig.~\ref{fig:loc-fam-phimean05} for fixed $\bar\phi=0.5$. The liquid state is linearly stable at all positive and small negative $r$ and becomes linearly unstable at $r\approx-0.75$. The primary bifurcation where the crystalline state emerges is again supercritical followed by a subcritical bifurcation where, as above, two branches of symmetric localized states emerge. These are again connected by branches of asymmetric localized states. In contrast to Fig.~\ref{fig:loc-fam-rm09}, here the LS$_\mathrm{odd}$ and LS$_\mathrm{even}$ branches no longer terminate on the branch of periodic states but continue towards lower $r$. However, the snaking stops after a finite number of saddle-node bifurcations, i.e., the localized structure does not fill up the entire domain. Note, that we do not consider here top-hat structures that involve plateaus at concentrations $\psi\neq0$ that are also present in the Swift-Hohenberg equation \citep{OuFu1996pre,BuKn2006pre,BuKn2007pla}. Such states can occur for large negative $r$ at appropriate $\bar\phi$.

\begin{figure}
		\centering
	(a)\includegraphics[width=0.6\hsize]{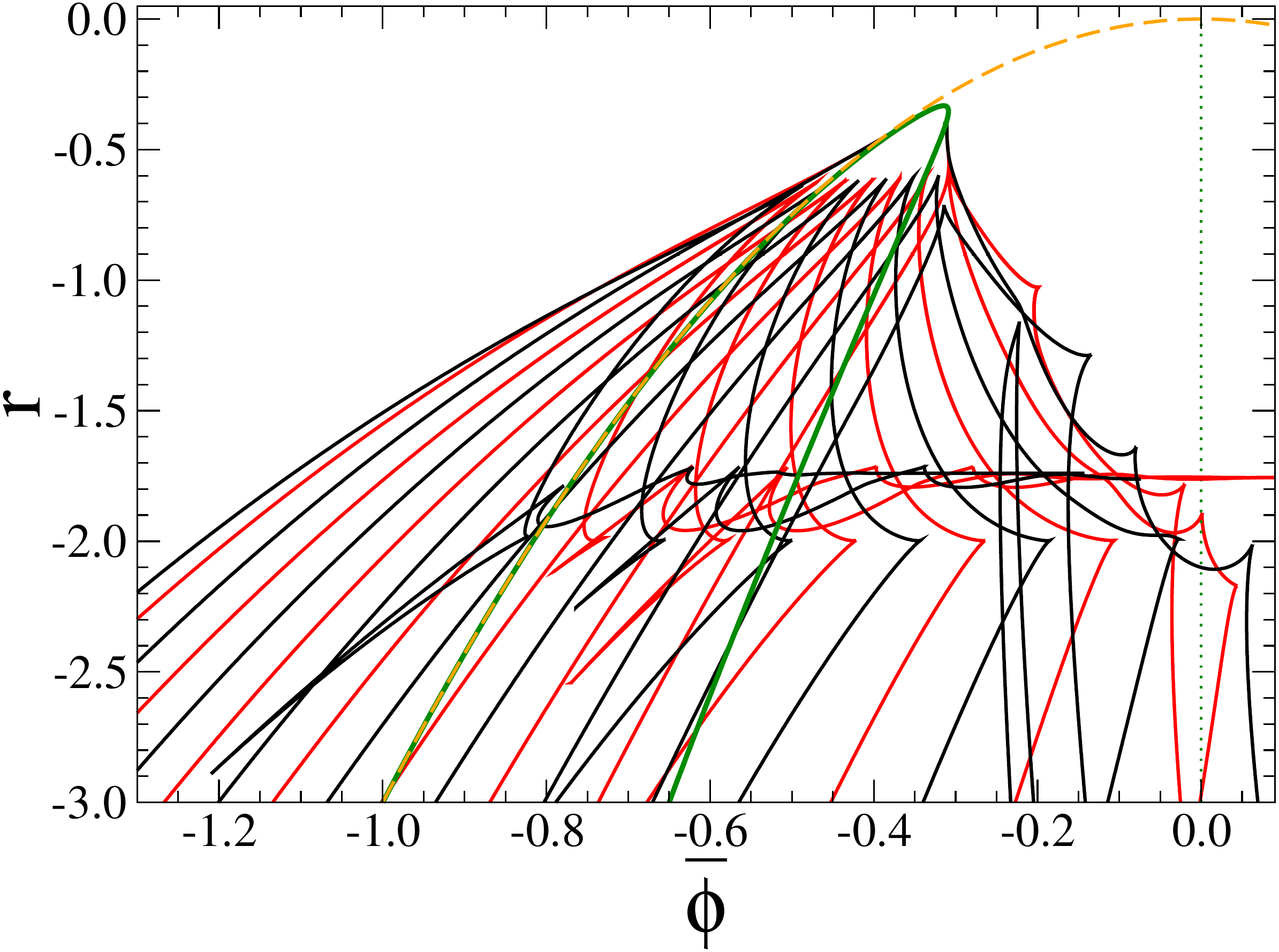}\\
	(b)\includegraphics[width=0.6\hsize]{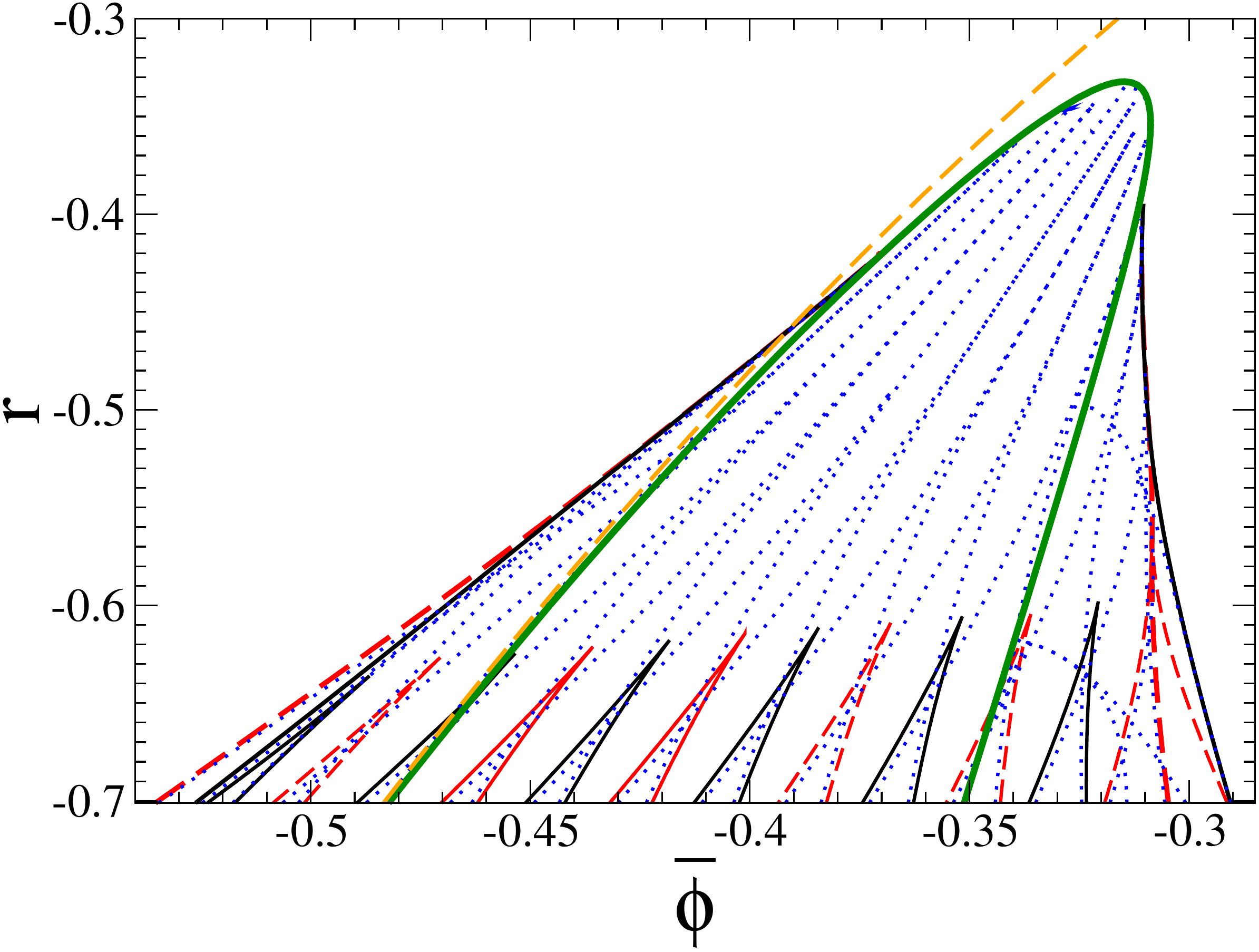}  
	\caption{(a) The loci of all saddle-node bifurcations of the branches of symmetric localized states visible in Figs.~\ref{fig:loc-fam-rm09} and~\ref{fig:loc-fam-phimean05} in the $(\bar\phi,r)$-plane for $r>-3.0$. (b) Magnification of the region of (a) close to the tricritical point including as blue dotted lines the loci of all pitchfork bifurcations where branches of asymmetric localized states emerge from branches of symmetric localized states. In both panels, the solid green line tracks the secondary bifurcations where symmetric localized states emerge from the periodic state, while the orange dashed line shows the locus of the primary bifurcation. Remaining parameters and line colors correspond to Fig.~\ref{fig:loc-fam-rm09}.
	}
	\mylab{fig:prof-loc-folds}
\end{figure}

The snaking behavior in both Fig.~\ref{fig:loc-fam-rm09} and Fig.~\ref{fig:loc-fam-phimean05} is summarized by tracking (using two-parameter continuation) the loci of all saddle-node bifurcations of localized states in the parameter plane spanned by the mean density and effective temperature, displayed in Fig.~\ref{fig:prof-loc-folds}. Also shown are loci of primary and secondary pitchfork bifurcations where periodic and localized states emerge, respectively. Figure~\ref{fig:prof-loc-folds}(a) presents the results for all saddle-node bifurcations of the symmetric localized states visible in Figs.~\ref{fig:loc-fam-rm09} and~\ref{fig:loc-fam-phimean05}. The accompanying panel~(b) provides a magnification of the range $-1<r<-0.35$ and additionally includes the loci of all pitchfork bifurcations where branches of asymmetric localized states emerge. Both panels show the loci of primary and secondary bifurcations where the periodic states bifurcate from the uniform state and the symmetric localized states emerge from the periodic ones, respectively.

From Fig.~\ref{fig:prof-loc-folds} we see that for increasing $r$ the saddle-node bifurcations annihilate pairwise in hysteresis bifurcations (visible as cusps). This happens first at small $\bar\phi$ and successively for larger $\bar\phi$, alternating between the LS$_\mathrm{odd}$ and LS$_\mathrm{even}$ branches. Above the cusps the branches of localized states still wiggle, but saddle-node bifurcations are absent. However, the branches of rung states remain and ultimately vanish one by one through collision with the secondary bifurcation [Fig.~\ref{fig:prof-loc-folds}(b)]. Note that adjacent secondary bifurcations at low $r$ approach one another and annihilate at $r\approx-0.33$ where localized states cease to exist \citep{TARG2013pre}. The thermodynamic tricritical point where the first order phase transition becomes a second order transition is at slightly larger $r=-9/38\approx-0.24$ (not shown here, but see \citet{TARG2013pre}).

\begin{figure}
		\centering
	\includegraphics[width=0.8\hsize]{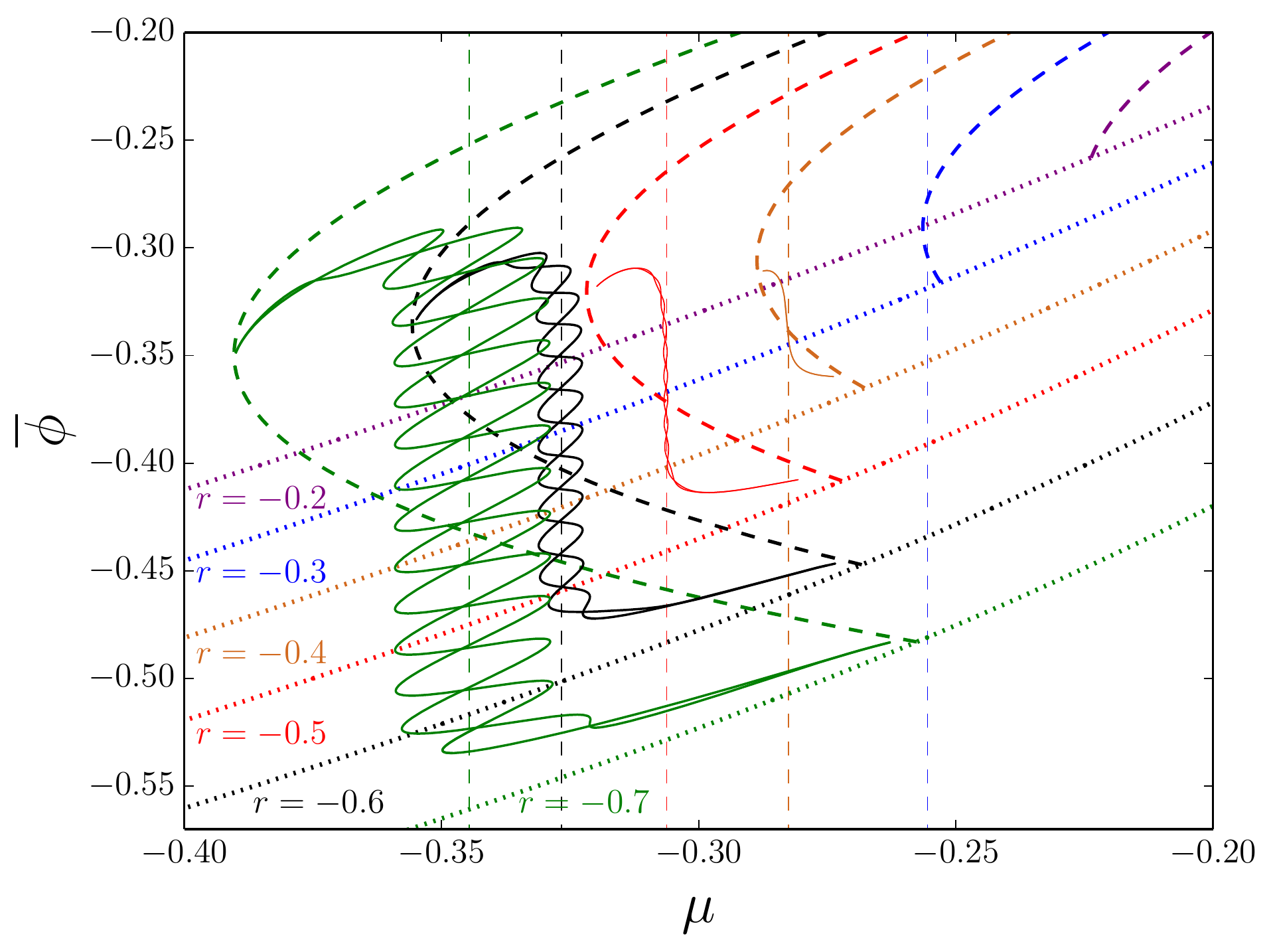}  
	\caption{Branches of liquid (dotted line), crystalline (dashed lines) and symmetric localized (thin solid lines) steady states for the conserved Swift-Hohenberg equation (\ref{eq:csh}) are given. Shown is the mean concentration $\bar\psi$ as a function of the chemical potential $\mu=\tilde\mu+(1+r)\bar\phi$. Various values of $r$ are used as indicated. The vertical dashed lines indicate the $\mu$-values at liquid-crystal coexistence (Maxwell construction).
		Remaining parameters are as in Fig.~\ref{fig:loc-fam-rm09}.
	}
	\mylab{fig:bif-over-mu}
\end{figure}

Interestingly, the tricritical point does not correspond to the point where the primary bifurcation in Fig.~\ref{fig:loc-fam-rm09} and~\ref{fig:loc-fam-phimean05} changes its character from super- to subcritical as occurs in other systems, e.g., the standard Swift-Hohenberg equation. Such a transition also exists here, but occurs at a much more negative temperature, $r=-9/2$. This brings us to the final point of this section, namely, the question of how the steady states and resulting bifurcation diagrams of the standard (nonconserved) Swift-Hohenberg equation (\ref{eq:sh}) and of its conserved counterpart (\ref{eq:csh}) are related.  Note that the equation describing the steady states of the latter [Eq.~(\ref{eq:csh-loc-steady})] is identical to the equation for the steady states of the Swift-Hohenberg equation (\ref{eq:sh}) with an additional external field $\mu$ \citep{KoCh2006prl}.

To determine the bifurcation diagram of Fig.~\ref{fig:loc-fam-rm09}, one solves Eq.~(\ref{eq:csh-loc-steady}) using $\bar\phi$ as a control parameter. As the dynamics (\ref{eq:csh}) conserves $\bar\phi$, the bifurcations in Fig.~\ref{fig:loc-fam-rm09} correctly indicate the linear stability of states under the conserved dynamics. The allowed perturbations are then at fixed $\bar\phi$. However, as we control $\bar\phi$, along the solution branches the chemical potential $\mu=\tilde\mu+(1+r)\bar\phi$ (Lagrange multiplier) has to be adapted. This allows us to present the identical curves using $\mu$ as the main control parameter, as shown for a number of $r$-values in Fig.~\ref{fig:bif-over-mu}.
Remarkably, in the representation as function of  $\mu$ the slanted snaking of Fig.~\ref{fig:loc-fam-rm09} becomes vertically aligned snaking centered around the coexistence chemical potential values (as obtained in a Maxwell construction). Although this alternative representation has ``straightened'' the slanted snaking into aligned snaking it invalidated the diagram as a bifurcation diagram for steady states of the conserved SH dynamics. Instead, as $\mu$ changes along the solution branches, the bifurcations in Fig.~\ref{fig:bif-over-mu} indicate the linear stability of states under nonconserved dynamics, Eq.~(\ref{eq:sh}), with an imposed external field $\mu$. The allowed perturbations are then at fixed $\mu$ and allow for a variation in $\bar\phi$. In this case the transition between on the one hand first and second order phase transition and on the other hand subcritical and supercritical bifurcation coincide. For further discussion of this point see the conclusion of \citet{TARG2013pre} and~\citet{EGUW2019springer} where this issue is also discussed for nonconserved and conserved Allen-Cahn equations. The latter is also known as the Cahn-Hilliard equation.
%
\section{Traveling localized states in an active PFC model} \mylab{sec:apfc}
%
To obtain a simple dynamical continuum model for the crystallization processes of active colloids, the PFC equation (\ref{eq:csh}) is coupled with the dynamics of a polarization field $\mathbf{P}$ that indicates the local strength and direction of polar order and the related active drive. The polarization is a vectorial order parameter with a nonconserved dynamics. The resulting coupled system is the active PFC model \citep{MeLo2013prl}
\begin{align}
\partial_{t}\psi &=  \nabla^{2}\frac{\delta\mathcal{F}}{\delta\psi}-\alpha\nabla\cdot\mathbf{P},\nonumber\\
\partial_{t}\mathbf{P} &= \nabla^{2}\frac{\delta\mathcal{F}}{\delta\mathbf{P}}-D_{\mathrm{r}}\frac{\delta\mathcal{F}}{\delta\mathbf{P}}-\alpha\nabla\psi,
\label{eq:gov}
\end{align}
where $\alpha$ is the coupling strength, also called the activity parameter or velocity of self-propulsion. The polarization is subject to translational and rotational diffusion with $D_{\mathrm{r}}$ as the rotational diffusion constant. The energy functional
\begin{equation}
\mathcal{F}[\psi,\mathbf{P}]=\mathcal{F}_\mathrm{sh}[\psi]+\mathcal{F}_{\mathbf{P}}[\mathbf{P}]
\end{equation}
is the sum of the standard phase-field-crystal functional (\ref{eq:sh-en}) and an orientational part
\begin{equation}
\mathcal{F}_{\mathbf{P}}[\mathbf{P}]=\int \mathrm{d^n}{\bf r} \left(\tfrac{1}{2}c_1|\mathbf{P}|^{2}+\tfrac{1}{4}c_2|\mathbf{P}|^{4}\right).
\label{eq:functionalP}
\end{equation}

The uncoupled dynamics ($\alpha=0$) correspond to purely conserved and mixed nonconserved and conserved gradient dynamics on $\mathcal{F}[\psi,\mathbf{P}]$, respectively, with no coupling between density $\psi$ and polarization $\mathbf{P}$. The coupling terms cannot be written in gradient form, i.e., the coupling is purely nonvariational. Note that it has the simplest nontrivial form allowed by the respective scalar and vector character of the fields $\psi$ and $\mathbf{P}$ that maintains the conserved character of the $\psi$-dynamics.

Although, the functional $\mathcal{F}_{\mathbf{P}}$ with $c_1<0$ and $c_2>0$ allows for spontaneous polarization most works consider the simpler case $c_1>0$ and $c_2=0$ \citep{MeLo2013prl,MeOL2014pre,ChGT2016el,OpGT2018pre}. Then, spontaneous polarization is absent and polarization is solely created by density gradients while the variational part of its dynamics, i.e., diffusion, always acts as damping.
Introducing in this case the variations of Eqs.~(\ref{eq:sh-en}) and (\ref{eq:functionalP}) into the governing equations~(\ref{eq:gov}) one obtains the nondimensional evolution equations
\begin{align}
\partial_{t}\psi &= \nabla^{2}\left\{\left[r+\left(1+\nabla^{2}\right)^{2}\right]\psi+\left(\bar{\phi}+\psi\right)^{3}\right\}-\alpha\nabla\mathbf{\cdot P}, \label{eq:dtpsi} \\
\partial_{t}\mathbf{P} &= c_1\nabla^{2}\mathbf{P} - D_{\mathrm{r}}c_1\mathbf{P}-\alpha\nabla\psi. 
\label{eq:dtP}
\end{align}
In 1D the polarization $\vec{P}$ becomes a scalar $P$ that represents the strength and directional sense of motion. The solution behavior of Eqs.~(\ref{eq:dtpsi}) and~(\ref{eq:dtP}) is studied by \citet{MeLo2013prl,MeOL2014pre,ChGT2016el} and~\citet{OpGT2018pre} by means of linear stability analysis of the uniform state, direct time simulations in 1D and 2D and by numerical bifurcation analysis tracking various nonlinear states and their linear stability.

Equation (\ref{eq:dtpsi}), like Eq.~(\ref{eq:csh}), conserves $\psi$ and hence $\int_V\psi\,\mathrm{d^n}{\bf r}=0$. This in turns implies that in the long-time limit $\int_V\mathbf{P}\,\mathrm{d^n}{\bf r}=0$. However, in contrast to the passive PFC model reviewed in section~\ref{sec:pfc} that exhibits small-scale stationary linear instabilities and steady nonlinear states, the nonvariational character of the active model also allows for short-wave oscillatory linear instability modes of the uniform state, related primary Hopf bifurcations and drift bifurcations whereby traveling nonlinear states emerge from steady ones. As for the passive PFC model, in the active PFC model there exist domain-filling periodic states \citep{MeLo2013prl} as well as localized states \citep{OpGT2018pre} where finite patches of resting or traveling crystalline state coexist with a uniform background. However, as the model is nonvariational, the presence of these states cannot be related to a Maxwell construction in the thermodynamic limit as done for the passive case by \citet{TFEK2019njp}.

\begin{figure}
		\centering
	\includegraphics[width=1.0\hsize]{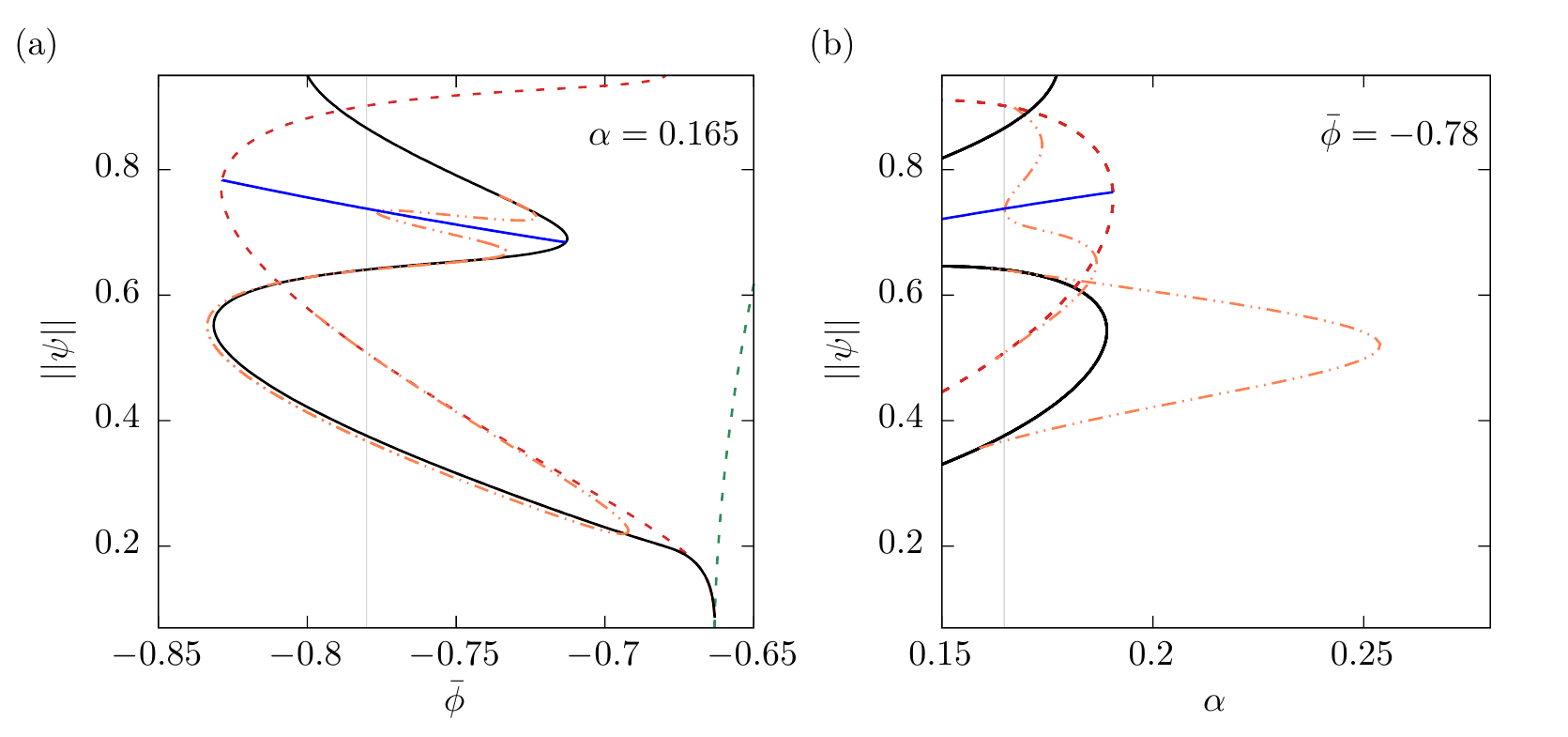}
	\caption{Typical bifurcation diagrams for the one-species active PFC model in 1D [Eqs.~(\ref{eq:dtpsi})-(\ref{eq:dtP})] focusing on branches of steady and traveling localized states using (a) the mean density $\bar\phi$ at fixed $\alpha=0.165$ and (b) the activity parameter $\alpha$ at fixed $\bar\phi=-0.78$ as main control parameters. Steady symmetric localized states with an even and an odd number of peaks are shown as dashed red and solid black lines, respectively. Also included are steady asymmetric localized states (blue lines), traveling localized states (dot-dashed lines), and periodic steady states with $n=16$ peaks (green dashed line). For clarity, only branches of one- and two-peak states are shown. Vertical gray lines indicate the value of the parameter fixed in the other panel. The remaining parameters are $r=-1.5$, $c_1=0.1$, $c_2=0$, and $D_r=0.5$. The domain size is $L=100$.}
	\label{fig:loc-apfc-bif-psibar}
\end{figure}

Figures~\ref{fig:loc-apfc-bif-psibar}(a) and~(b) present examples of typical bifurcation behavior of steady and traveling localized states employing the mean density $\bar\phi$ and the activity parameter $\alpha$ as main control parameters, respectively. In panel (a) one sees that steady localized states again exhibit slanted homoclinic snaking qualitatively similar to the passive case shown in Fig.~\ref{fig:loc-fam-rm09}. Only the first few wiggles are shown in order to reveal the finer details. Increasing $\bar\phi$, the uniform state becomes linearly unstable at $\bar\phi\approx-0.66$ where the $n=16$ periodic state bifurcates supercritically (green dashed line). The latter becomes unstable at small amplitude giving rise to two branches (black and red dashed) of steady symmetric localized states that emerge in a subcritical secondary pitchfork bifurcation just as in the passive case. The resulting branches of symmetric steady localized states with even and odd number of peaks are again connected by branches of asymmetric steady localized states (blue lines). Their existence as steady states is related to the linear coupling in Eqs.~(\ref{eq:dtpsi}) and (\ref{eq:dtP}) and the 'accidental' global conservation of $P$ mentioned above. Increasing $c_2$ from zero in (\ref{eq:functionalP}) or introducing a nonlinear coupling causes all rung states to drift. This will be further discussed elsewhere.

Qualitatively new in the active PFC model, as compared to in the passive PFC model, are the branches of traveling localized states (dot-dashed orange lines). These emerge from the branches of resting symmetric localized states in a drift-pitchfork bifurcation and may connect to branches of steady asymmetric localized states via drift-transcritical bifurcations. The resulting states drift with a constant velocity that depends on the control parameters.
A detailed discussion of the onset of motion (including an analytical condition) and typical profiles of resting and traveling states can be found in \citet{OpGT2018pre}. Note that Fig.~\ref{fig:loc-apfc-bif-psibar}(a) shows the case of relatively low $\alpha\approx0.16<\alpha_c$, close to but below the critical activity value $\alpha_c$ where the primary bifurcation becomes oscillatory. Slanted homoclinic snaking of traveling states at $\alpha>\alpha_c$ is also possible.

Figure~\ref{fig:loc-apfc-bif-psibar}(b) shows that an increase of the activity parameter $\alpha$ at fixed $\bar\phi$ may result in a transition from steady to traveling localized states. The $\alpha$ values for the onset of motion for the various states are very similar, but not identical. An interesting feature is the ``nose'' of traveling states that reaches far beyond the $\alpha$-range that allows for steady localized states. The locus of the saddle-node where it ends depends strongly on $\bar\phi$ and may even diverge to $\alpha\to\infty$ (cf.~Figs.~8 and~9 by \citet{OpGT2018pre}). The existence regions of the various localized states and the scattering behavior of traveling localized states is investigated by \citet{OKGT2020c}. There the occurrence of more involved localized states corresponding, e.g., to localized standing waves and traveling modulated waves is also discussed.

\begin{figure}
		\centering
	\includegraphics[width=0.8\hsize]{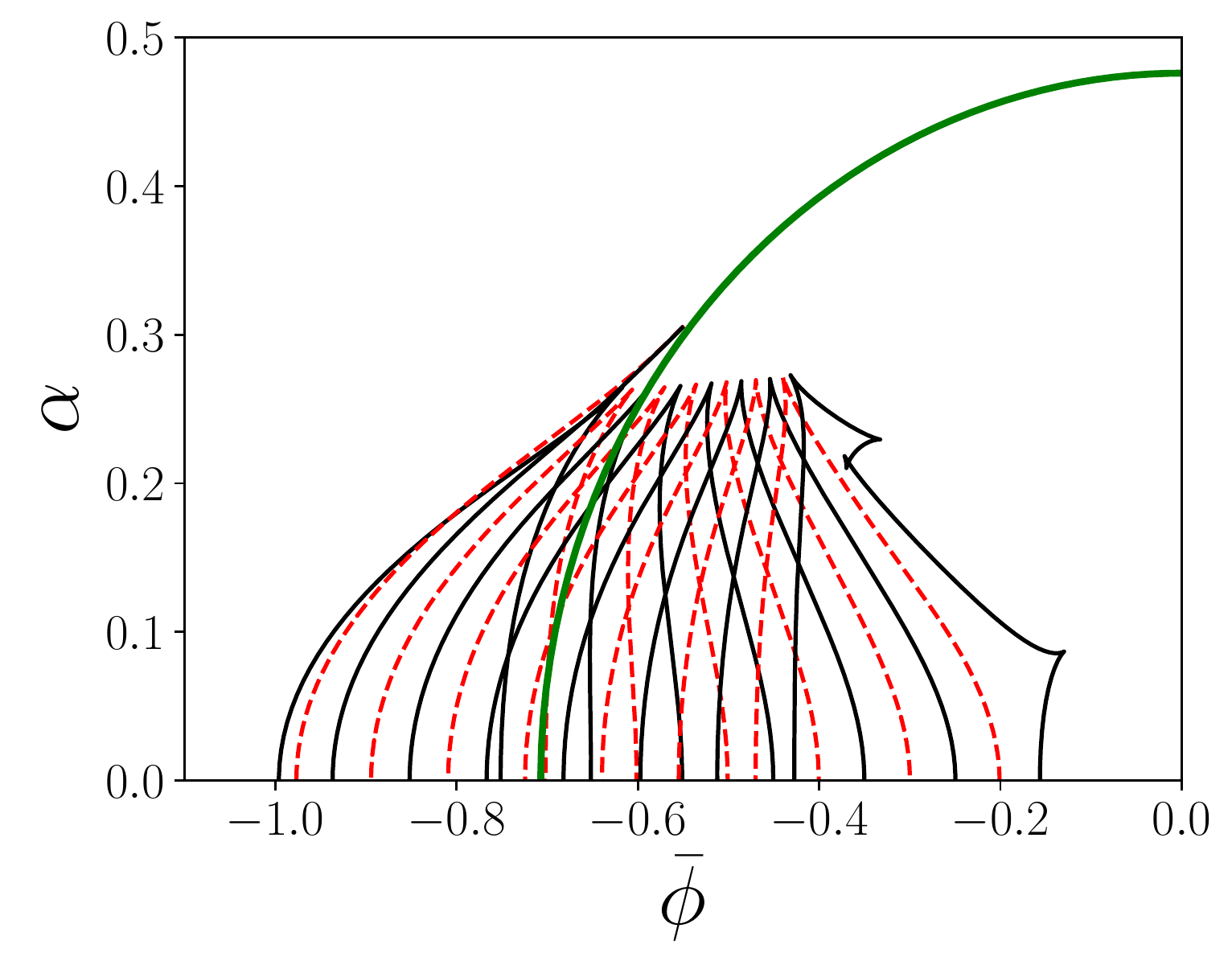}
	\caption{Loci of saddle-node bifurcations on the branches of resting symmetric localized states with even (dashed red lines) and odd (solid black lines) numbers of peaks in the $(\bar\phi,\alpha)$ plane. The locus of the primary bifurcation to periodic states is shown as a thick green line. The remaining parameters are as in Fig.~\ref{fig:loc-apfc-bif-psibar}.}
	\mylab{fig:apfc-folds}
\end{figure}

Finally, we discuss how the loci of the saddle-node bifurcations in Fig.~\ref{fig:loc-apfc-bif-psibar} depend on the other control parameter. In particular, Fig.~\ref{fig:apfc-folds} tracks their loci in the $(\bar\phi,\alpha)$-plane. It is remarkable that the overall structure is very similar to Fig.~\ref{fig:prof-loc-folds} with the activity in Fig.~\ref{fig:apfc-folds} taking the role of the effective temperature in Fig.~\ref{fig:prof-loc-folds}. 
This observation merits further investigation.
%
\section{Two coupled PFC models} \mylab{sec:2pfc}
%
A relatively simple continuum model for the crystallization dynamics of a mixture of two species of colloidal particles is obtained by coupling two PFC equations.  To study the behavior of a thermodynamic system the coupling is taken to be variational, while in the study of active colloids a nonvariational coupling may also be considered. Here, we focus on a linear coupling containing variational and nonvariational contributions of strength $c$ and $\alpha$, respectively. The resulting equations
\begin{align}
\partial_t \psi_1 = \partial_x^2[r\psi_1 + (\partial_x^2 + q_1^2)^2\psi_1 + (\psi_1 + \bar{\phi}_1)^3 + (c + \alpha)\psi_2],\notag\\
\partial_t \psi_2 = \partial_x^2[r\psi_2 + (\partial_x^2 + q_2^2)^2\psi_2 + (\psi_2 + \bar{\phi}_2)^3 + (c - \alpha)\psi_1]\label{eq:cPFC}
\end{align}
are parity-symmetric and also have the inversion symmetry $ \bar{\phi}_j+\psi_j \rightarrow - (\bar{\phi}_j+\psi_j )$. The dynamics conserve both mean densities $\bar{\phi}_j$, i.e., the $\psi_j$ again represent deviations from the mean with $\int_V {\mathrm d}x \, \psi_j =0$. For $ \alpha=0 $ the model represents conserved gradient dynamics on the energy functional
\begin{eqnarray}
\mathcal{F}[\phi_1,\phi_2] = \mathcal{F}_\mathrm{sh}[\phi_1] + \mathcal{F}_\mathrm{sh}[\phi_2] + \mathcal{F}_\mathrm{int}[\phi_1,\phi_2]
\mylab{eq:sh2-en}
\end{eqnarray}
with $\mathcal{F}_\mathrm{int}[\phi_1,\phi_2]= \int \mathrm{d^n}{\bf r} c \phi_1\phi_2$.
In principle, the two species of colloidal particles do not need to have the same typical length scale $ q_j $ nor the same coefficients of the gradient energy or local energy terms. However, here we keep things simple and consider only the case $ q_1 = q_2 = 1 $. We also use the same coefficients in all the other terms.

In section~\ref{sec:2pfc-vari} we focus on the fully variational case, but in the subsections that follow we explore two ways of making the model active. In particular, section~\ref{sec:2pfc-vari} considers Eqs.~\eqref{eq:cPFC} with $\alpha\neq0$ while \ref{sec:2pfc-nonvari} uses $ \alpha=0 $ but couples one of the densities to a polarization in a similar fashion as in section \ref{sec:apfc}. 

From here on, the domain size is $ L = 16\pi $, corresponding to 8 peaks at the onset of instability. The smaller domain results in bifurcation diagrams that are less crowded and easier to read. We expect qualitatively similar behavior for larger domains.
\subsection{Variational coupling}
\label{sec:2pfc-vari}
We set $ \alpha = 0 $ in Eq.~\eqref{eq:cPFC} to retain the variational character of the coupled PFC equations. In this case only resting states occur.
%


%

\begin{figure}[tbh]
		\centering
	\includegraphics[width=0.7\hsize]{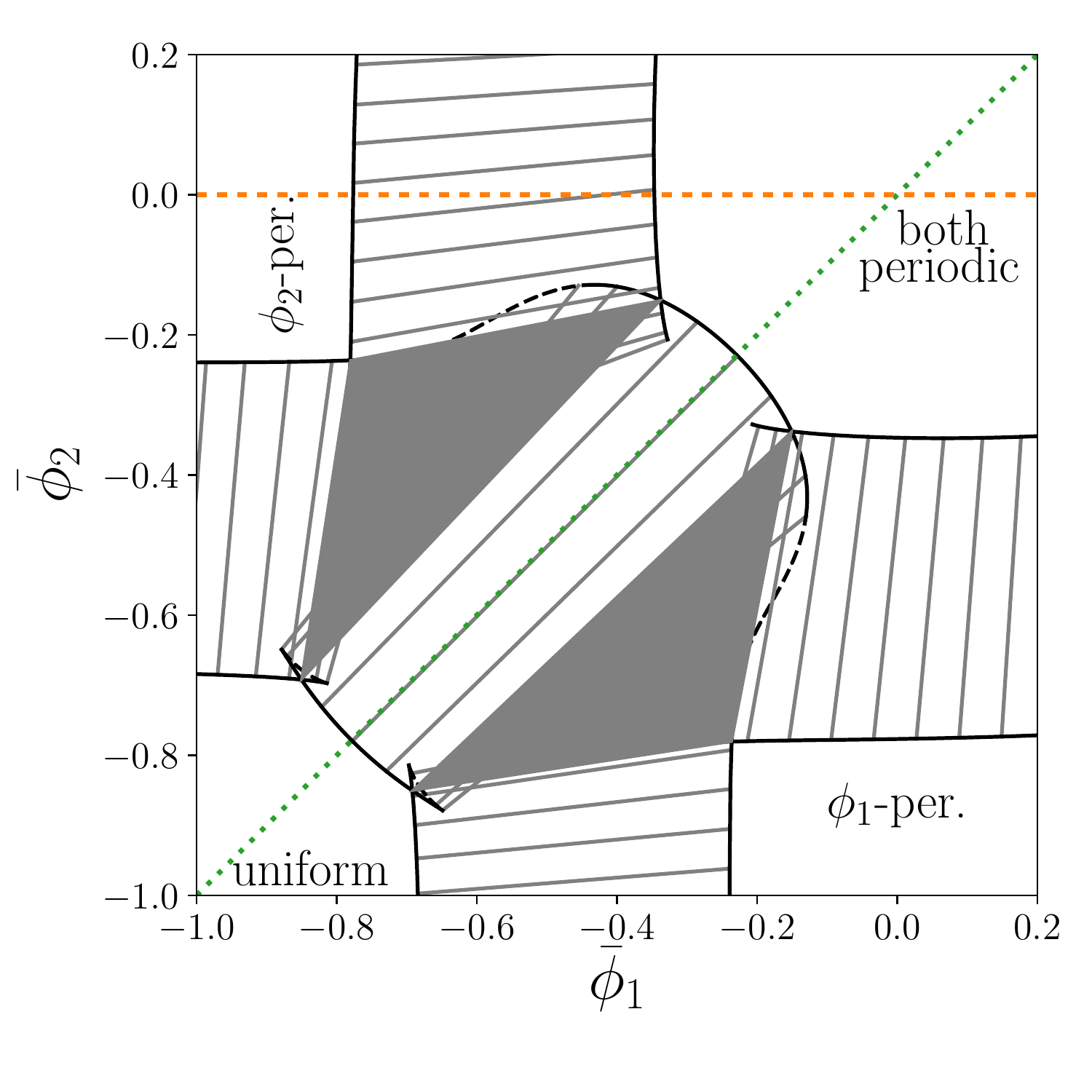}
	\caption{Phase diagram for the two-species passive PFC model  in 1D [Eq.~(\ref{eq:cPFC})] with $\alpha=0$ in the plane spanned by the mean densities $\bar\phi_{1}$ and $ \bar{\phi}_2$. The hatched  [gray shaded] areas represent regions of coexistence of two [three] phases, and the grey tie lines connect particular coexisting states on pairs of binodal lines. The green dotted and orange dashed straight lines correspond to the paths taken in the bifurcation diagrams of Figs.~\ref{fig:pfc2field_liquid-alloy} and \ref{fig:pfc2field_periodic-alloy}, respectively. The remaining parameters are $ r = -0.9 $ and $ c = -0.2 $.}
	\label{fig:pfc2field_phase_diagram}
\end{figure}

The passive system describes the crystallization behavior of a mixture of two species of colloid allowing for various liquid and crystalline phases. \citet{HoAT2020jpcm} employ continuation techniques to study in detail the thermodynamic phase behavior of the model in one and two dimensions. Figure~\ref{fig:pfc2field_phase_diagram} presents one quadrant of a typical phase diagram in the $(\bar\phi_1,\bar\phi_2$)-plane in 1D. The phase plane features four distinct phases: a low density uniform (well-mixed liquid) phase, and three different periodic (crystal) phases. The latter are a $\phi_1$-crystal in which the amplitude of the $\phi_1$ modulation is much larger than that in $\phi_2$, the corresponding $\phi_2$-crystal, and a crystalline ``alloy'' phase where the amplitude of the periodic modulation in both fields is similar. 

\begin{figure}[tbh]
		\centering
	\includegraphics[width=0.8\hsize]{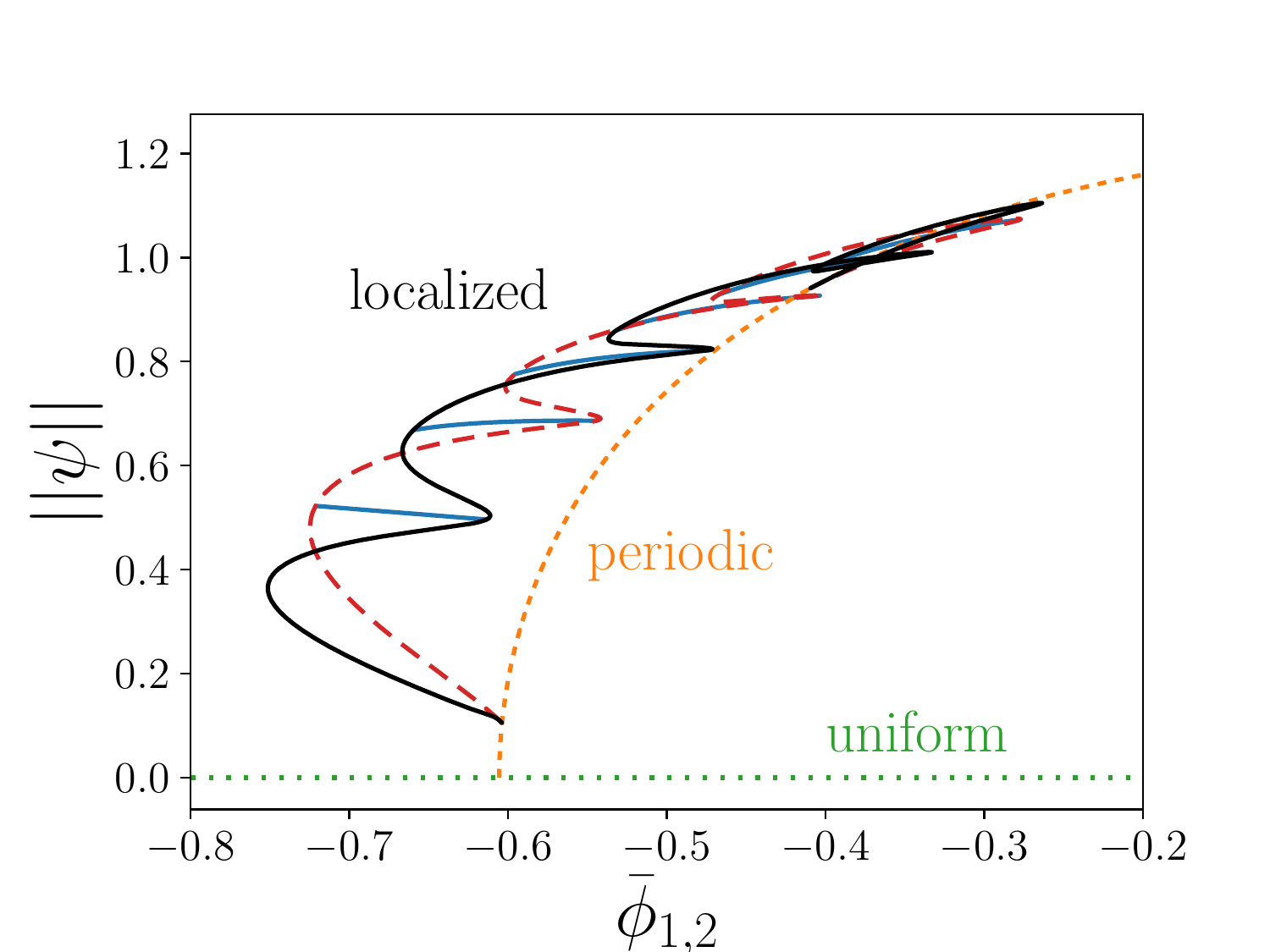}
	\caption{Typical bifurcation diagram using the mean densities $\bar\phi=\bar\phi_{1}=\bar\phi_{2}$ as a control parameter, crossing the uniform liquid state into the crystalline alloy coexistence region, i.e., corresponding to the dotted diagonal line in Fig.~\ref{fig:pfc2field_phase_diagram}. Steady symmetric localized states with even and odd numbers of peaks are shown as dashed red and solid black lines, respectively. Also included are steady asymmetric localized states (blue lines), the trivial uniform (liquid) state (green dotted line), and the periodic steady states with $n=8$ peaks (orange dashed line). The remaining parameters are as in Fig.~\ref{fig:pfc2field_phase_diagram}.}
	\label{fig:pfc2field_liquid-alloy}
\end{figure}

\begin{figure}[tbh]
		\centering
	\includegraphics[width=0.7\hsize]{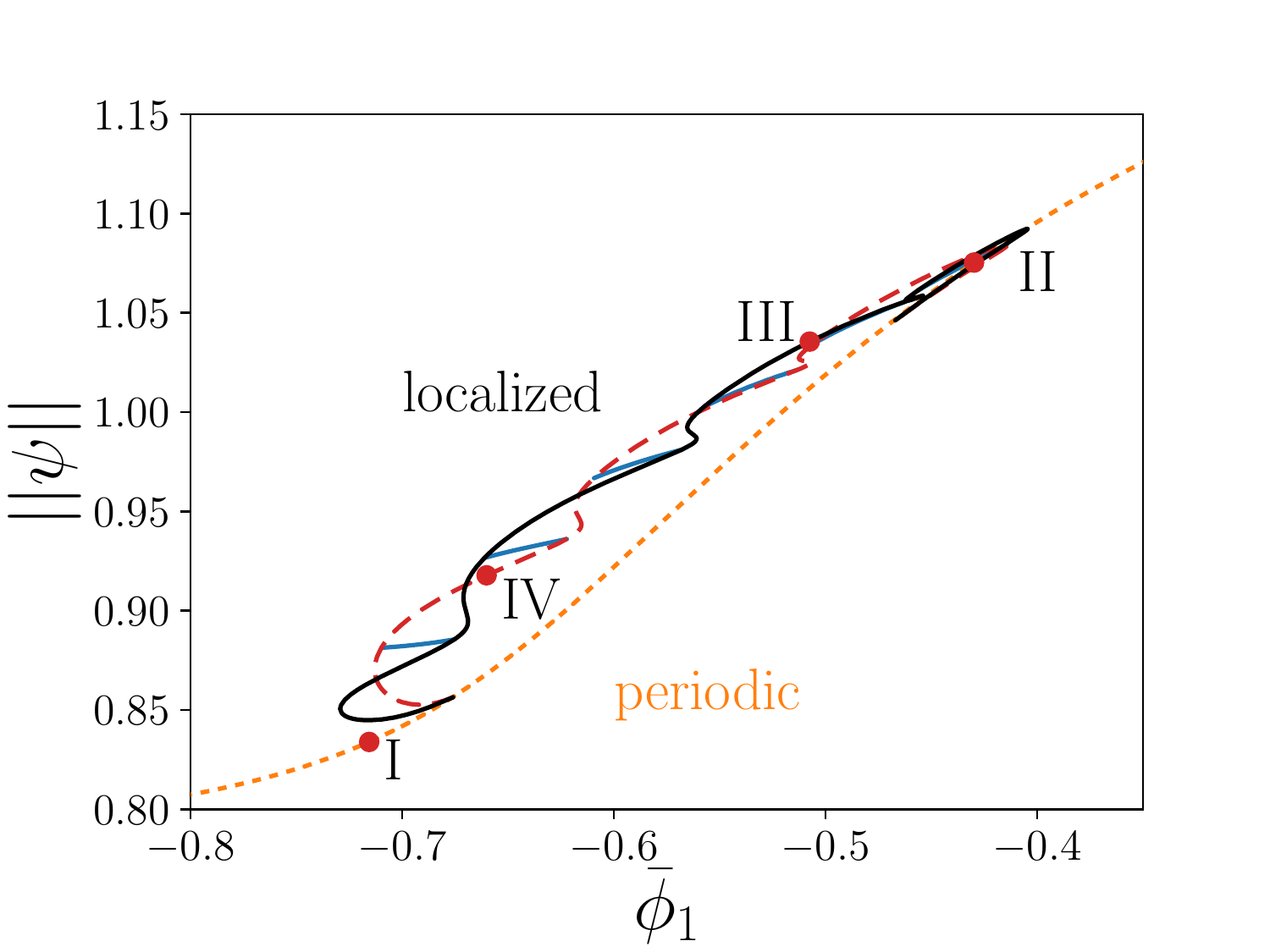}\\
	\includegraphics[width=0.7\hsize]{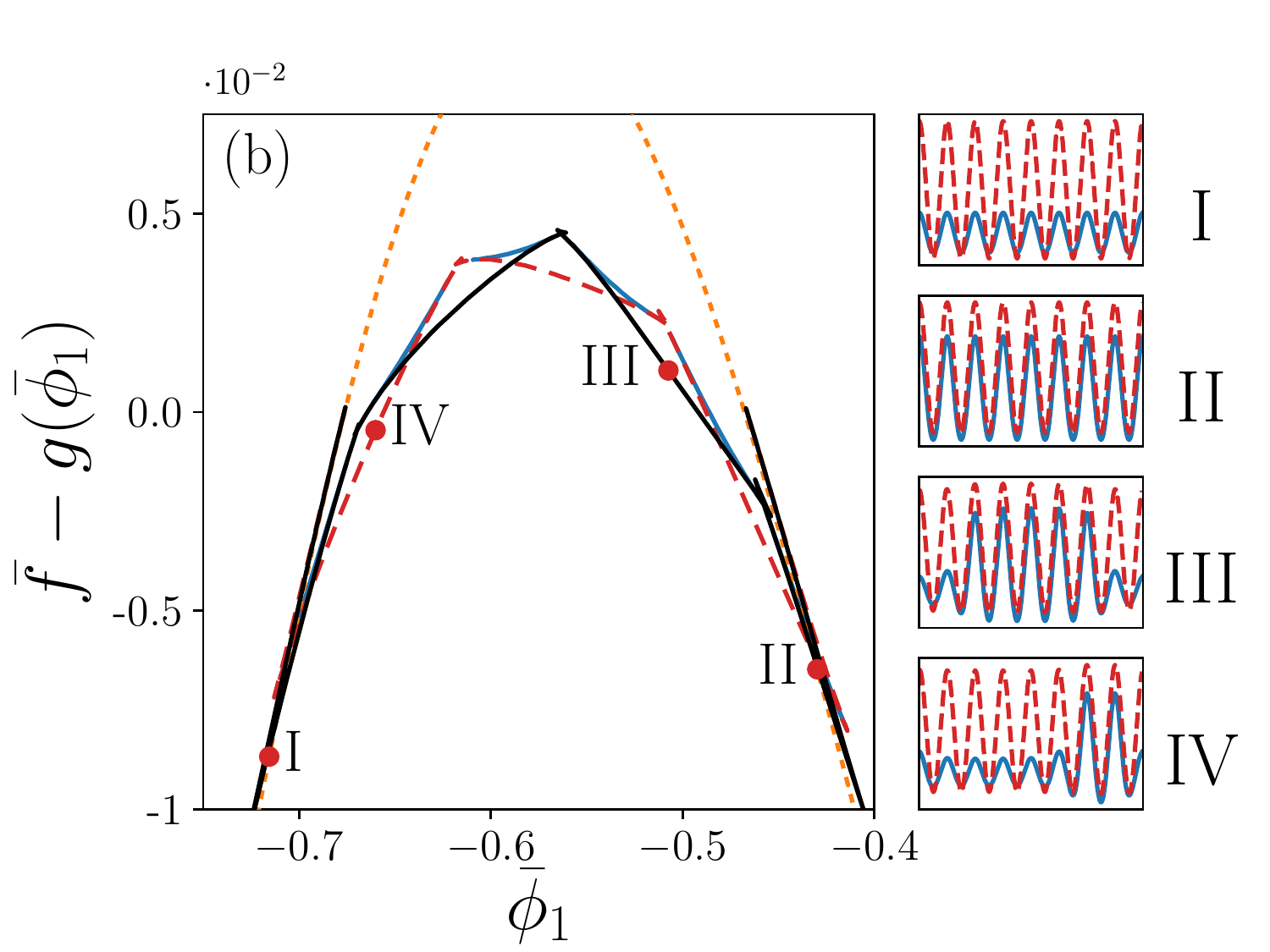}
	\caption{(a) Typical bifurcation diagrams using the mean density $\bar\phi_{1}$ as the control parameter at fixed $ \bar{\phi}_2 = 0 $ and crossing the $\phi_2$-crystal to crystalline alloy coexistence region, i.e., corresponding to the dashed horizontal line in Fig.~\ref{fig:pfc2field_phase_diagram}. Line styles are as in Fig.~\ref{fig:pfc2field_liquid-alloy} while the remaining parameters are as in Fig.~\ref{fig:pfc2field_phase_diagram}. Panel (b) presents the corresponding free energy densities $f$, showing that localized states are the lowest energy states in the coexistence region. The function $ g(\bar{\phi}_1) $ is subtracted for clarity and corresponds to a straight line connecting the bifurcation points where the localized states emerge and terminate. The small panels I to IV give a selection of concentration profiles on the various branches as indicated by the filled circles in (a) and (b).}
	\label{fig:pfc2field_periodic-alloy}
\end{figure}

The boundaries between the various phases in Fig.~\ref{fig:pfc2field_phase_diagram} are formed by two-phase coexistence regions indicated by pairs of binodal lines and three-phase coexistence regions. Particular coexisting states on the binodal lines are connected by straight tie lines. The continuation of binodal lines beyond the corners of the shaded triangles corresponds to metastable (solid lines) and unstable (dashed lines) coexistence. Note that the three-phase coexistence region corresponds to a triple point. For further details see \citet{HoAT2020jpcm}. The straight dotted and dashed lines show the parameter paths we take to determine bifurcation diagrams related to a liquid-crystal and a crystal-crystal phase transition, presented in Figs.~\ref{fig:pfc2field_liquid-alloy} and \ref{fig:pfc2field_periodic-alloy}, respectively.

Figure~\ref{fig:pfc2field_liquid-alloy} follows the path along the diagonal using $\bar\phi=\bar\phi_{1}=\bar\phi_{2}$ as the control parameter, i.e., from the liquid to the crystalline alloy phase border. As there is complete symmetry between the two fields, the bifurcation diagram is rather similar to Fig.~\ref{fig:loc-fam-rm09}. The uniform state becomes linearly unstable at $ \bar{\phi} \approx -0.61$ where the branch of periodic states with $ n = 8 $ peaks emerges in a supercritical pitchfork bifurcation. These states are initially linearly stable but lose stability in a secondary pitchfork bifurcation where branches of odd (LS$ _\text{odd} $) and even (LS$ _\text{even} $) symmetric localized states bifurcate subcritically as in section \ref{sec:pfc}. Six rungs of steady asymmetric localized states (blue curves) connect the two branches of symmetric states. When the domain is filled the branches of localized states terminate back on the $ n=8 $ periodic state. The slanted snaking structure is less dense than in Fig.~\ref{fig:loc-fam-rm09} since the domain is smaller.

In contrast, Fig.~\ref{fig:pfc2field_periodic-alloy}(a) follows a path using $ \bar{\phi}_1 $ as the control parameter at fixed $ \bar{\phi}_2 = 0 $, passing the  phase boundary from the $\phi_2$-crystal to the crystalline alloy. In the transition region localized states in $ \phi_1 $ exist on a background of a domain-filling periodic state in $ \phi_2 $. Starting at small values of $ \bar{\phi}_1 $ states on the branch of periodic states have a large amplitude in $ \phi_2 $ and a small amplitude in $ \phi_1 $. When increasing $ \bar{\phi}_1 $, the periodic state becomes unstable at $ \phi_1\approx -0.68 $ where two branches of localized states emerge at a subcritical pitchfork bifurcation. The LS$ _\text{odd} $ state is again stabilized when it folds back in a saddle-node bifurcation, while the LS$ _\text{even} $ branch is stabilized after a saddle-node and a subsequent pitchfork bifurcation, where the first rung branch of steady asymmetric localized states emerges. In the same manner as before, the localized states form a strongly slanted snakes-and-ladders structure along which more and more density peaks are added at the edges of the localized states with a strong focus on the $\phi_1$ field. When peaks of both fields fill the domain the snaking branches terminate on the periodic branch in another subcritical bifurcation. There, the periodic branch regains its linear stability and features crystalline alloy states, i.e., the modulations in both fields are of similar large amplitude.

Interestingly, it is not possible to spot the $\phi_2$-crystal to crystalline alloy first order phase transition by just considering the branch of periodic states as it is smooth and shows no special features in the representation of Fig.~\ref{fig:pfc2field_periodic-alloy}(a). Indeed, it is the existence of localized states as global energy minima [see Fig.~\ref{fig:pfc2field_periodic-alloy}(b)] that indicates the phase transition. The phase transition can most easily be spotted when plotting the mean grand potential over the appropriate chemical potential, as it is then related to a swallow tail structure (not shown here, but \citep[cf.][]{HoAT2020jpcm}).

To summarize, the two bifurcation diagrams presented in Figs.~\ref{fig:pfc2field_liquid-alloy} and \ref{fig:pfc2field_periodic-alloy} consider finite size systems on two paths across regions where first order phase transitions occur. In Fig.~\ref{fig:pfc2field_liquid-alloy} the background state is of a different symmetry than the pattern within the localized state while in Fig.~\ref{fig:pfc2field_periodic-alloy} this is not the case. 
This difference matters when approaching the thermodynamic limit: while in Fig.~\ref{fig:pfc2field_liquid-alloy} the secondary bifurcation where the localized states emerge approaches the primary bifurcation as the domain size increases; in contrast, this does not happen for the case in Fig.~\ref{fig:pfc2field_periodic-alloy}.
In the present variational case we do not consider how the loci of saddle-node bifurcations change with coupling strength $c$ or effective temperature $r$ but expect the behavior to be similar to Fig.~\ref{fig:loc-fam-rm09}.

\subsection{Nonvariational coupling}
\label{sec:2pfc-nonvari}
\begin{figure}[tbh]
		\centering
	\includegraphics[width=0.8\hsize]{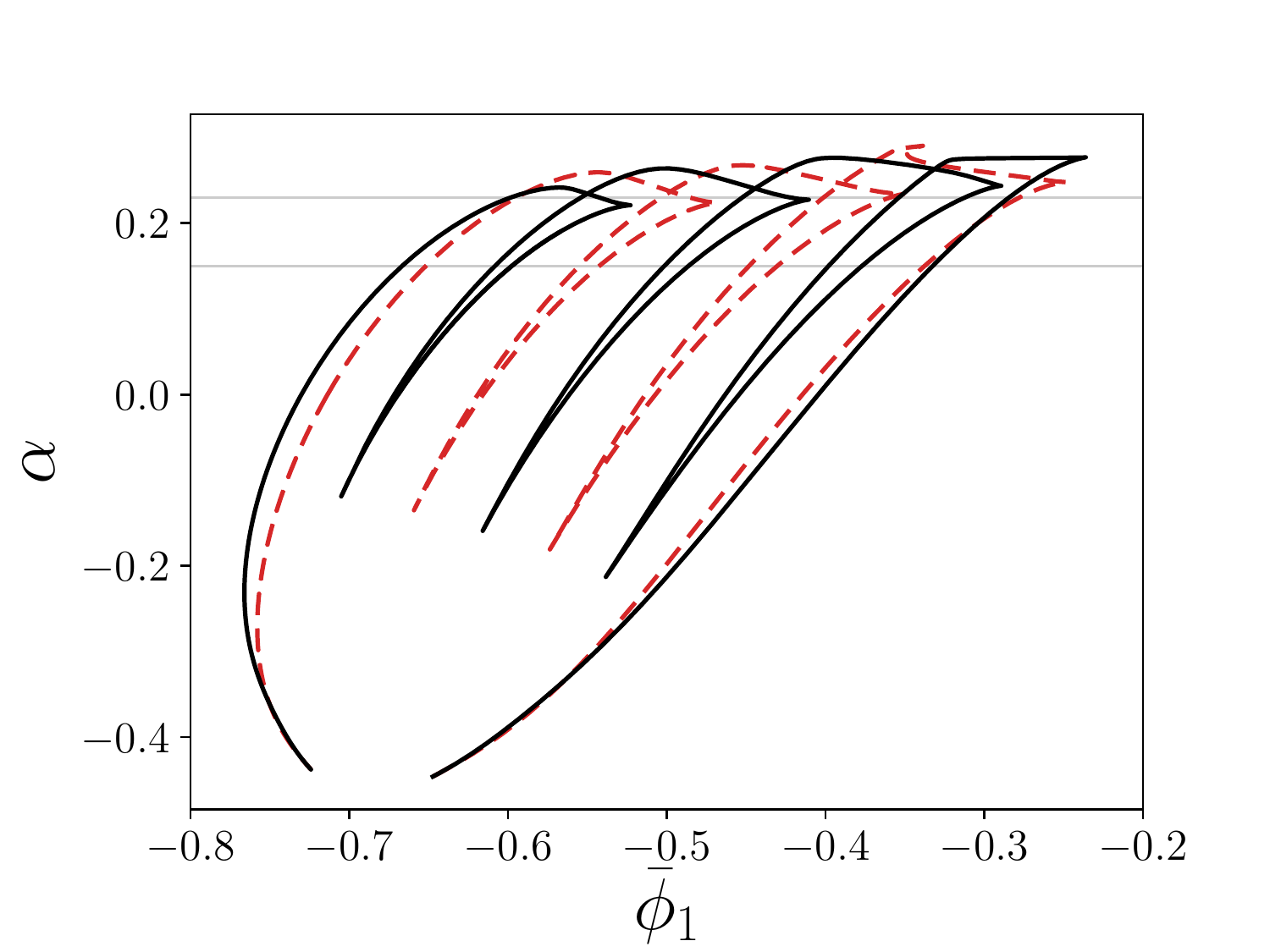}
	\caption{Characterization of localized states in the nonvariational two-field PFC model, Eq.~\eqref{eq:cPFC} with $ \alpha\neq 0$. Shown are the loci of all saddle-node bifurcations that are present at $ \alpha = 0 $ on the branches of steady symmetric localized states in the $(\bar\phi_1,\alpha)$-plane. The dashed red and solid black lines indicate states with even and odd peak numbers, respectively. The thin gray horizontal lines mark the paths taken in the bifurcation diagrams in Fig.~\ref{fig:pfc2field-nonvar-bif}. The remaining parameters are as in Fig.~\ref{fig:pfc2field_periodic-alloy}.}
	\label{fig:pfc2field_folds_epsilon-psibar}
\end{figure}

Next we briefly consider the case of nonvariational coupling, i.e., Eq.~\eqref{eq:cPFC} with $ \alpha\neq 0$. To directly obtain an overview of the influence of the nonvariational coupling strength $\alpha$ on the snaking behavior, we consider how the loci of all saddle-node bifurcations in Fig.~\ref{fig:pfc2field_periodic-alloy} change with $\alpha$. The result is shown in Fig.~\ref{fig:pfc2field_folds_epsilon-psibar}. We note that with increasing $|\alpha|$, saddle-node bifurcations approach each other in the $(\bar\phi_1,\alpha)$-plane and eventually annihilate with one another. However, this occurs in a different manner for positive and negative $\alpha$. For decreasing $ \alpha<0 $ the behavior is quite similar to what we see in the passive one-field PFC for increasing temperature $r$ [cf.~Fig.~\ref{fig:prof-loc-folds}(b)] or in the active one-field PFC for increasing activity [cf.~Fig.~\ref{fig:apfc-folds}]. Namely, the saddle-node bifurcations disappear pairwise in cusp-like structures within the range $\alpha\approx-0.12\dots-0.22$. As a result those of the more extended localized states disappear last. Finally, the outer saddle-node bifurcations disappear when the secondary bifurcations become supercitical, at about $\alpha\approx-0.44$. As in the models considered previously, a ``wiggling'' slanted structure without saddle-nodes persists and corresponds to localized states that continuously increase in size along the branch. The localized state branches ultimately disappear at slightly smaller $\alpha$, but the periodic states continue to exist.

\begin{figure}[tbh]
		\centering
	\includegraphics[width=0.75\hsize]{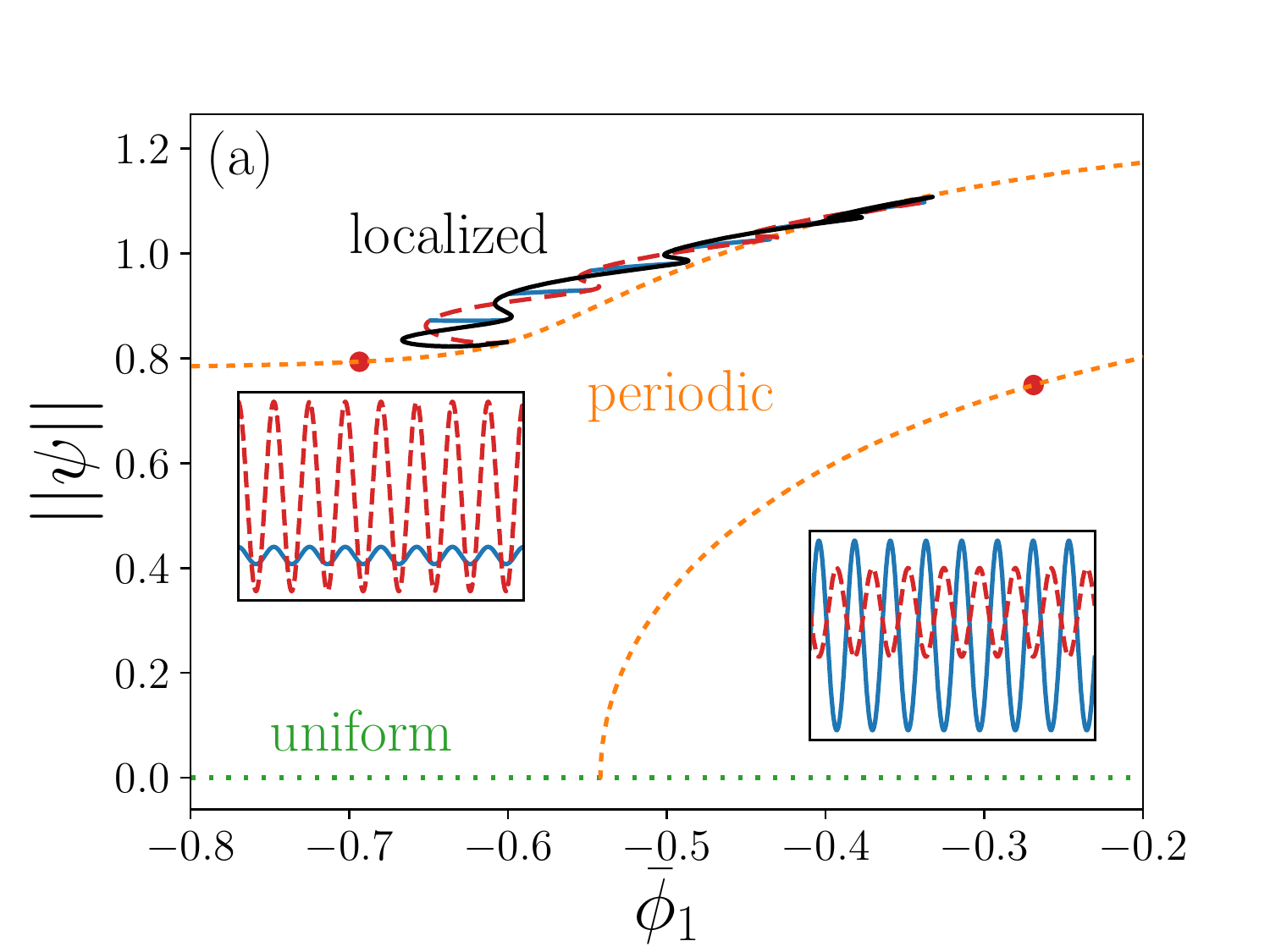}\\
	\includegraphics[width=0.75\hsize]{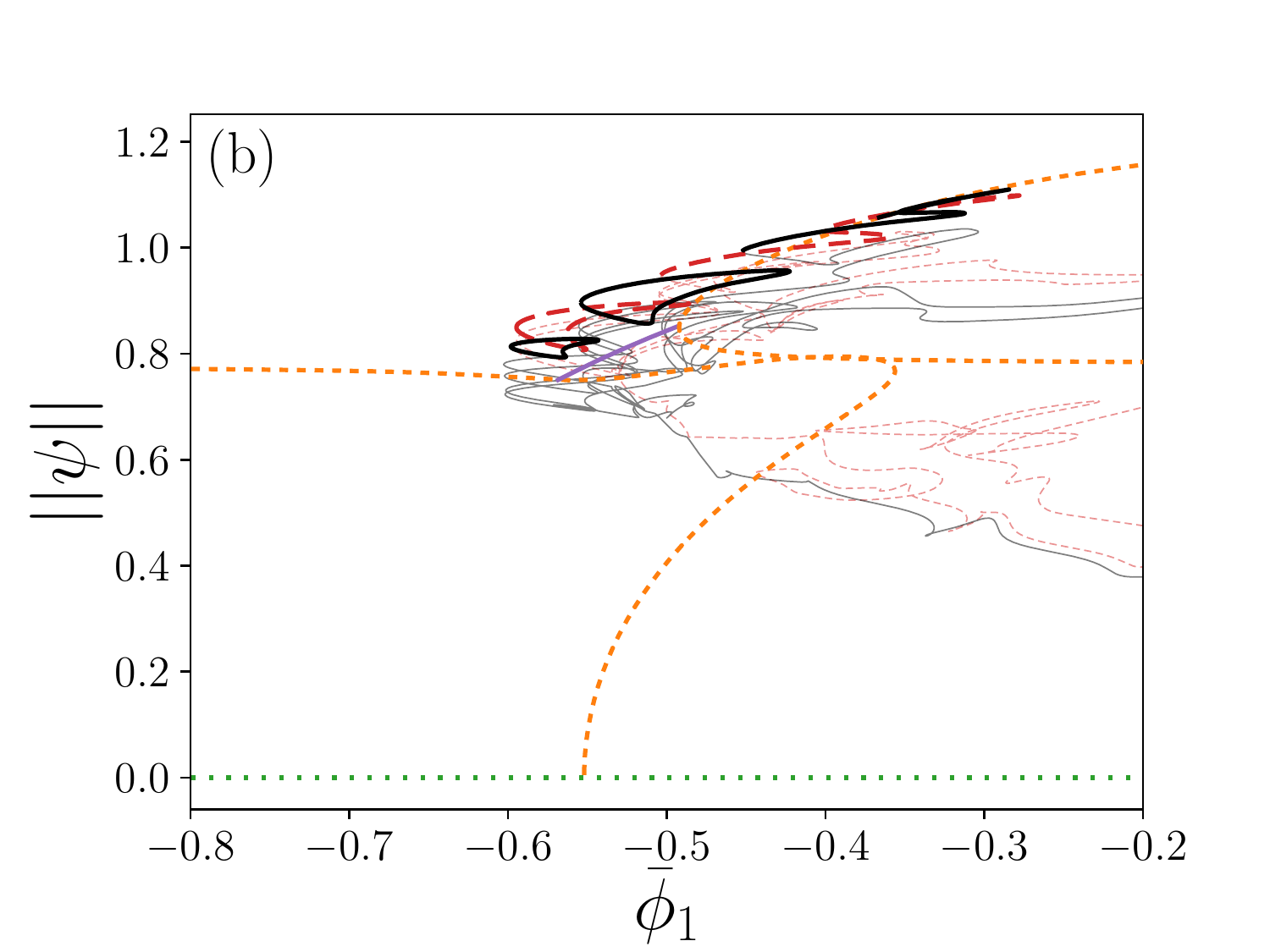}
	\caption{Typical bifurcation diagrams for the nonvariationally coupled two-field PFC model. The mean density $\bar\phi_{1}$ is used as a control parameter at fixed $ \bar{\phi}_2 = 0 $ for (a) $\alpha=0.15$ and (b) $\alpha=0.23$, i.e., corresponding to cuts through Fig.~\ref{fig:pfc2field_folds_epsilon-psibar} along the horizontal gray lines. Line styles are as in Fig.~\ref{fig:pfc2field_liquid-alloy} while the remaining parameters are as in Fig.~\ref{fig:pfc2field_folds_epsilon-psibar}. (a) Red dots mark the loci of the sample profiles in the insets. (b) The thick purple line corresponds to traveling periodic states. Thin gray [pink] lines represent further branches of mostly unstable LS$ _\text{odd} $ [LS$ _\text{even} $] states beyond the last saddle-node bifurcation known from Fig.~\ref{fig:pfc2field_folds_epsilon-psibar}.}
	\label{fig:pfc2field-nonvar-bif}
\end{figure}

In contrast, for increasing $ \alpha>0 $ the pairs of saddle-node bifurcations that annihilate for negative $\alpha$ now drift apart. Shortly above $\alpha=0.2$ the right folds annihilate with the next-nearest left folds on the left, created at a lower value of $ \bar{\phi}_1$, while the nearest left folds annihilate with the next-nearest right folds on the right, created at a higher value of $ \bar{\phi}_1$ (see Fig.~\ref{fig:pfc2field_folds_epsilon-psibar}). This complex structure implies that a major reorganisation of the bifurcation diagram at constant $\alpha$ must occur. This is shown in Fig.~\ref{fig:pfc2field-nonvar-bif}. Panel~(a) presents a cut through Fig.~\ref{fig:pfc2field_folds_epsilon-psibar} at $\alpha=0.15$ below the region where the cusps occur. The branches of in-phase modulation of the two densities (that is stable at large and small $\bar\phi_1$) and of anti-phase modulation are shown as a function of $ \bar{\phi}_1$ with one sample profile each. The behavior is very similar to the passive case in Fig.~\ref{fig:pfc2field_periodic-alloy}. In particular, the asymmetric rung states are again steady as in the active PFC model in section~\ref{sec:apfc}, a fact that is again related to the linear coupling in Eqs.~(\ref{eq:cPFC}) as explained elsewhere.

Figure~\ref{fig:pfc2field-nonvar-bif}(b) follows a cut through Fig.~\ref{fig:pfc2field_folds_epsilon-psibar} at $\alpha=0.23$, i.e., after the first three right-pointing cusps have already occurred. We see that the bifurcation behavior is tremendously more involved than in Fig.~\ref{fig:pfc2field-nonvar-bif}~(a). When the variational and nonvariational coupling parameters have the same absolute strength, i.e., $ \alpha = -c $, the two branches of periodic states meet in a codimension-2 bifurcation transcritical bifurcation and reconnect in the opposite sense, i.e., at larger $ \alpha $ the branches with anti-phase modulation continue towards smaller $ \bar{\phi}_1 $, while the branch of in-phase modulation forms an isola at small $ |\bar{\phi}_1| $. The branch of anti-phase modulations is stable at low $ \bar{\phi}_1 $ until it is destabilized at $ \bar\phi_1 \approx -0.58 $. A second bifurcation on that branch, at $ \bar\phi_1 \approx -0.57 $, generates a branch of traveling periodic states. 
The isola of periodic in-phase modulation is comprised of stable solutions for $ |\bar{\phi}_1| < 0.37 $ where both fields show large amplitude modulation, i.e., at large L$^2 $ norm. At the bifurcation at $ \bar{\phi}_1 \approx - 0.37$, the branch of in-phase modulation is destabilized and, as in Figs.~\ref{fig:pfc2field-nonvar-bif}(a) and \ref{fig:pfc2field_periodic-alloy}, two branches of localized states emerge. At $ \alpha = 0.23 $ these two branches undergo the last four saddle-node bifurcations that were also present at $ \alpha = 0 $, before undergoing a complex series of bifurcations that we do not study in detail. In Fig.~\ref{fig:pfc2field-nonvar-bif}(b) the branches of localized states beyond these four saddle-node bifurcation are shown in light gray [pink].  As can be seen in Fig.~\ref{fig:pfc2field_folds_epsilon-psibar} the remaining saddle-node bifurcations present at $ \alpha = 0 $ are still present at $ \alpha= 0.23 $, even though the corresponding parts of the snaking branches are no longer connected to the periodic branches. At $ \alpha = 0.23 $, two right-pointing cusps involving states with odd peak numbers and one involving states with an even number of peaks have already occurred. Correspondingly, in Fig.~\ref{fig:pfc2field-nonvar-bif}(b) we find three isolas of localized states, one with an even number of peaks and two with an odd number of peaks. This indicates that at each of the right-pointing cusps in Fig.~\ref{fig:pfc2field_folds_epsilon-psibar} one of the isola buds off from the snaking branch. Even though the snaking branches are broken up, we can still recover the main features of snaking behavior. 

It is important to note that while the bifurcation diagram in Fig.~\ref{fig:pfc2field-nonvar-bif}(b) looks very complicated, at every value of $ \bar{\psi}_1 $ there still exists at least one simple linearly stable steady state, i.e., a state that is known from the passive case $ \alpha = 0 $. In addition, there are also branches of traveling localized states at the parameter values of Fig.~\ref{fig:pfc2field-nonvar-bif}(b). For clarity, they are omitted from the bifurcation diagram.

\section{Coupling active and passive PFC models} \mylab{sec:pfcapfc}
%
Finally, we look at a mixture of passive and active colloids by coupling a passive PFC model as in section~\ref{sec:pfc} and an active PFC model as in section~\ref{sec:apfc}. The model is
\begin{align}
\partial_t \psi_1 &= \partial_x^2[r\psi_1 + (\partial_x^2 + 1)^2\psi_1 + (\psi_1 + \bar{\phi}_1)^3 + c\psi_2] -\alpha \nabla\cdot\mathbf{P},\nonumber\\
\partial_t \psi_2 &= \partial_x^2[r\psi_2 + (\partial_x^2 + 1)^2\psi_2 + (\psi_2 + \bar{\phi}_2)^3 + c\psi_1],\nonumber\\
\partial_t \mathbf{P} &= c_1\nabla^2\mathbf{P} - D_rc_1\mathbf{P} - \alpha\nabla\psi_1,
\label{eq:appfc}
\end{align}
i.e., we combine a variational linear coupling (via the coupling strength $c$) between two densities $\psi_1$ and $\psi_2$ as in section~\ref{sec:2pfc-vari} with a coupling of the active density $\psi_1$ to the polarization field (via the coupling strength $\alpha$) in a similar manner as in section~\ref{sec:apfc}. The density $\psi_2$ represents a passive field and does not contribute to the polarization.

\begin{figure}[tbh]
		\centering
	\includegraphics[width=0.7\hsize]{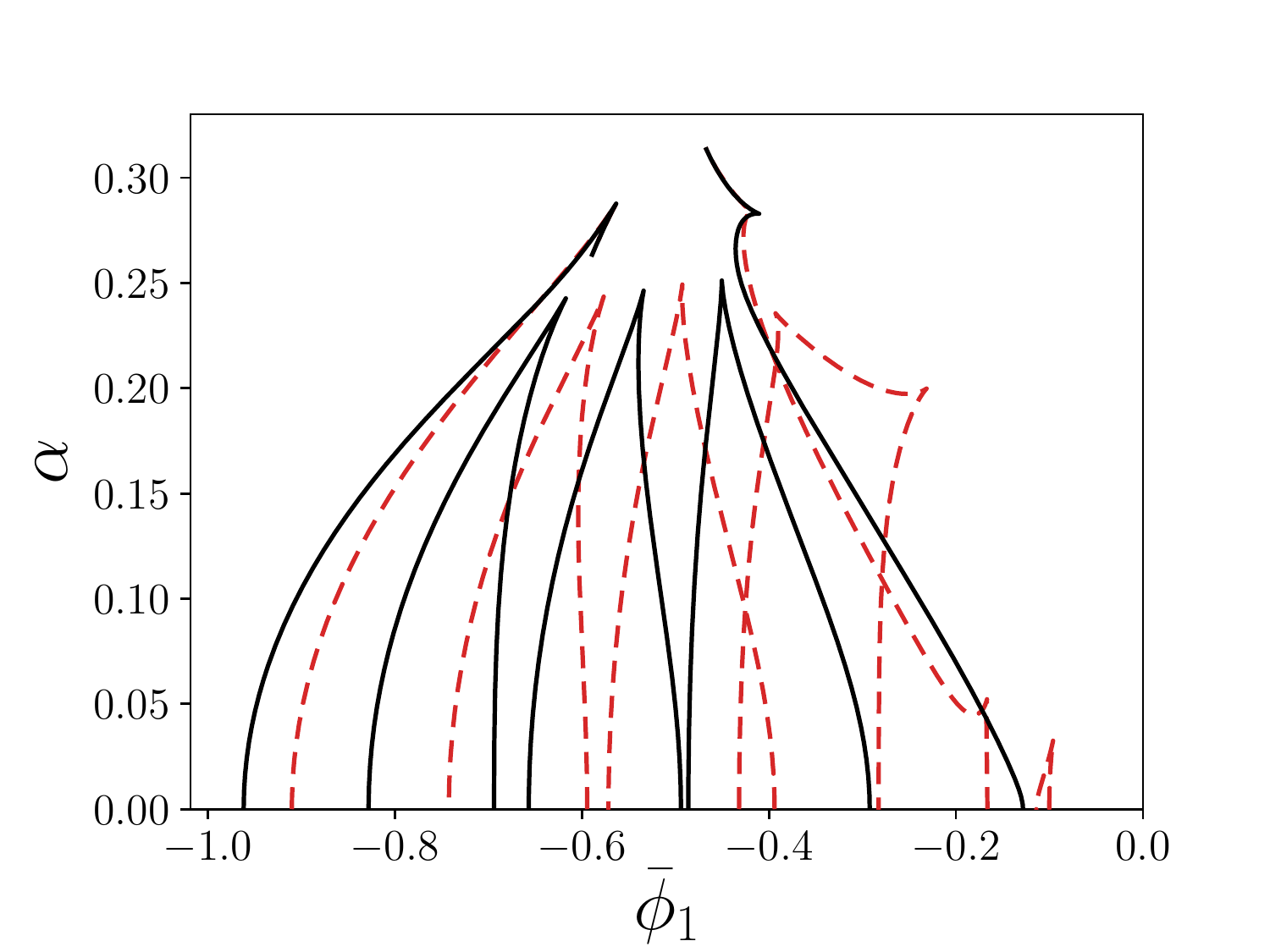}
	\caption{For the coupled active and passive PFC model [Eqs.~\eqref{eq:appfc}] we show the loci in the $(\bar\phi_{1},\alpha)$-plane of saddle-node bifurcations on the branches of resting symmetric localized states with even (dashed red lines) and odd (solid black lines) numbers of peaks. The remaining parameters are $ r = -1.5 $, $ \bar{\phi}_2 = -2 $ and $ c = -0.2 $.}
	\label{fig:apfc2field_folds_v0-psibar}
\end{figure}

\begin{figure}[tbh]
		\centering
	\includegraphics[width=0.7\hsize]{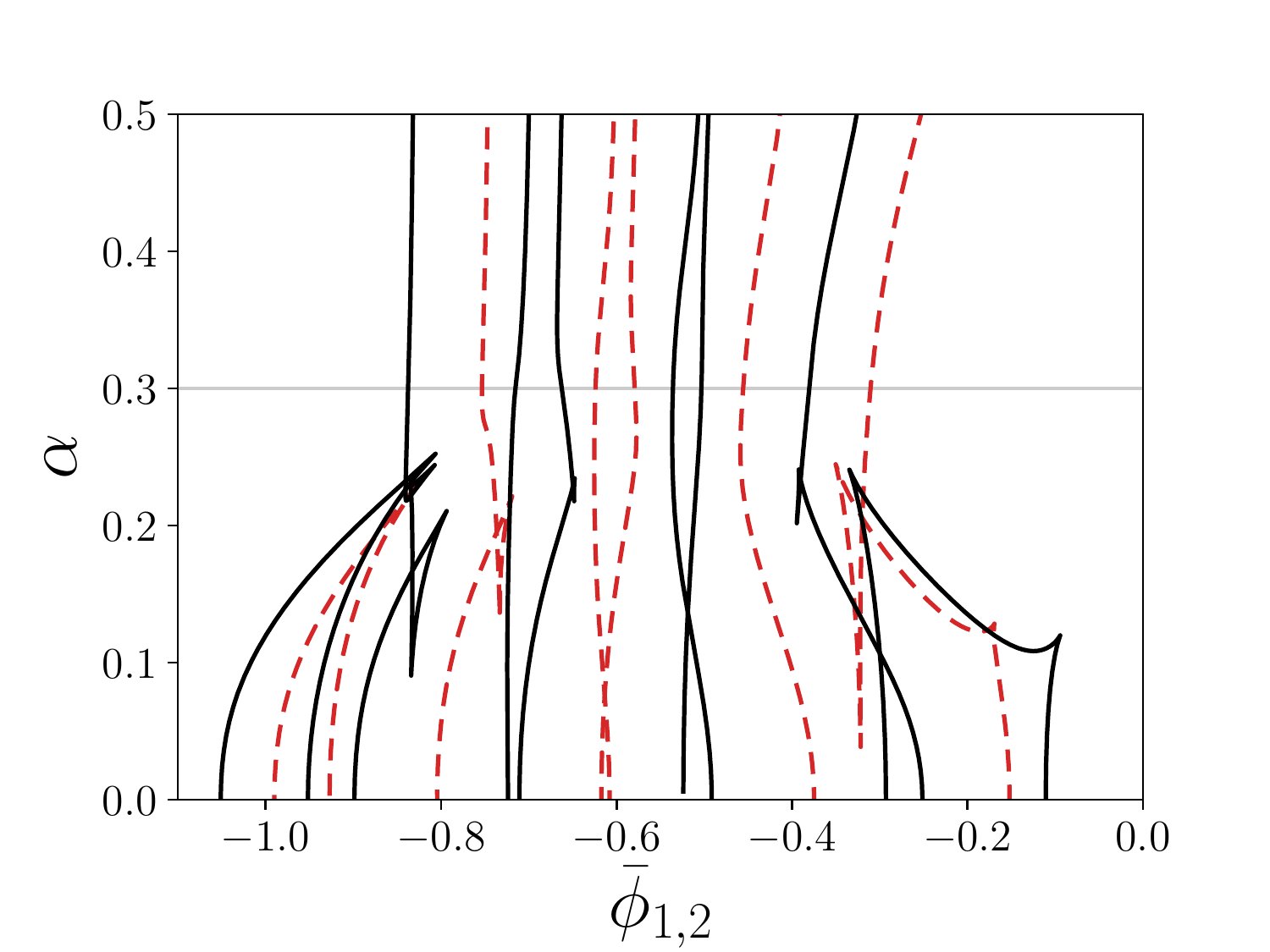}
	\caption{For the coupled active and passive PFC model [Eqs.~\eqref{eq:appfc}] we show loci in the $(\bar\phi,\alpha)$-plane of saddle-node bifurcations on the branches of resting symmetric localized states with even (dashed red lines) and odd (solid black lines) numbers of peaks where $\bar\phi=\bar\phi_{1}=\bar\phi_{2}$. Here $ r = -1.5 $ and the remaining parameters are as in Fig.~\ref{fig:pfc2field_liquid-alloy}. The horizontal line indicates the value of $\alpha$ for the bifurcation diagram in Fig.~\ref{fig:appfc-nonvar-bif}.}
	\label{fig:apfc2field_folds_v0-psibar_diagonal}
\end{figure}

As before, we get an initial overview of the snaking behavior with changing $ \alpha $ by tracking the loci of the saddle-node bifurcations at constant $ \bar\phi_2 = -2 $ in the $(\bar\phi_1,\alpha)$-plane (see Fig.~\ref{fig:apfc2field_folds_v0-psibar}) and along the diagonal $\bar\phi = \bar{\phi}_1 = \bar{\phi}_2 $ in the $(\bar\phi,\alpha)$-plane (see Fig.~\ref{fig:apfc2field_folds_v0-psibar_diagonal}). As in section \ref{sec:apfc} we choose $ r = -1.5$. 

We focus first on the case in Fig.~\ref{fig:apfc2field_folds_v0-psibar}. At the low mean density $ \bar\phi_2 = -2 $ the field $\phi_2$ is suppressed and has no major influence on the behavior of the system. We therefore expect to find behavior similar to the active PFC model in section \ref{sec:apfc}. Indeed, Fig.~\ref{fig:apfc2field_folds_v0-psibar} resembles Fig.~\ref{fig:apfc-folds}. The saddle-node bifurcations annihilate at $ \alpha \approx 0.25 $ and the activity again takes the role of the effective temperature in Fig.~\ref{fig:prof-loc-folds}. 

The loci of the saddle-node bifurcations along the diagonal $\bar{\phi} = \bar{\phi}_1 = \bar{\phi}_2 $ are presented in Fig.~\ref{fig:apfc2field_folds_v0-psibar_diagonal}. Here, only a minority of saddle-node bifurcations annihilate with increasing $\alpha$. The majority continue to exist for large $\alpha$ and their loci in Fig.~\ref{fig:apfc2field_folds_v0-psibar_diagonal} become nearly vertical. This can be understood by noting that $ \alpha $ takes on the role of a ``nonequilibrium effective temperature''. Since only $ \psi_1 $ describes the active colloid, only crystallites formed by $ \psi_1 $ start to melt as $\alpha$ increases. However, the crystalline structure in $ \psi_2 $ persists while the influence of both $ \psi_1 $ and (indirectly) $\mathbf{P} $ on these crystallites diminishes for large $\alpha$, i.e., $ \psi_1, P \to 0$ as $\alpha$ increases. 

\begin{figure}[tbh]
		\centering
	\includegraphics[width=0.7\hsize]{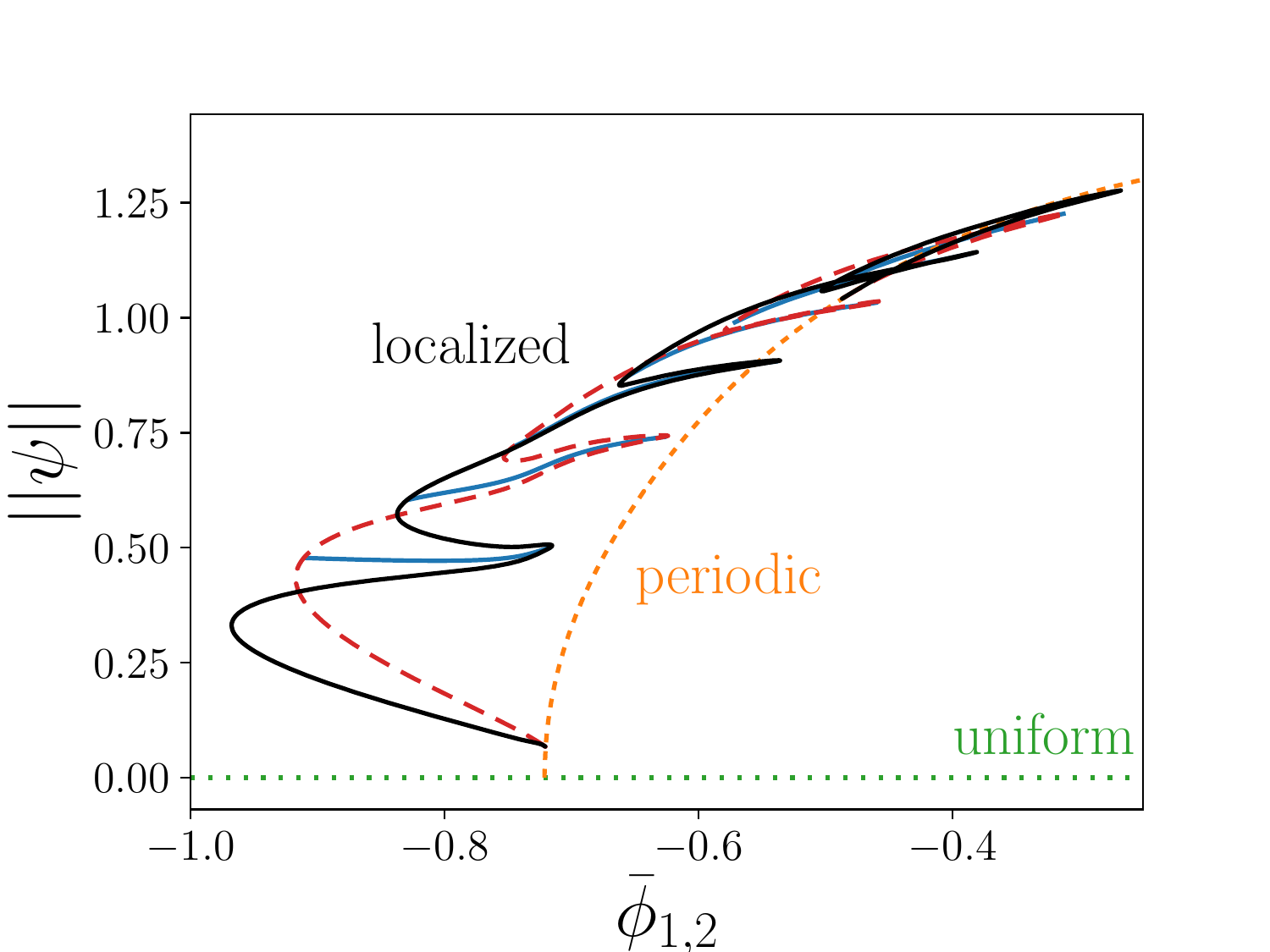}
	\caption{Typical bifurcation diagram for the coupled active and passive PFC model corresponding to the horizontal cut at fixed $\alpha=0.3$ through Fig.~\ref{fig:apfc2field_folds_v0-psibar_diagonal} as indicated by a horizontal line. Line styles are as in Fig.~\ref{fig:pfc2field_liquid-alloy} while the remaining parameters are as in Fig.~\ref{fig:apfc2field_folds_v0-psibar_diagonal}.}
	\label{fig:appfc-nonvar-bif}
\end{figure}

This observation is further illustrated in Fig.~\ref{fig:appfc-nonvar-bif} where we present a bifurcation diagram along a horizontal cut through Fig.~\ref{fig:apfc2field_folds_v0-psibar_diagonal} at fixed $\alpha=0.3$, i.e., above all the cusps. There, the slanted snaking is still present, but for smaller $ |\bar{\phi}_{1,2}| $, all states are destabilized by many Hopf bifurcations. The branches emerging there are not studied here, but promise very rich behavior that is worth investigating.

\section{Conclusions}
\mylab{sec:conc}

We have explored the occurrence of slanted homoclinic snaking of localized states in a number of passive and active phase-field crystal (PFC) models. The standard one-field PFC model \citep{ElGr2004pre,ELWG2012ap} corresponds to the conserved Swift-Hohenberg equation and is perhaps the simplest example of a pattern-forming system with a conserved quantity. The bifurcation structure involving snaking branches of localized states is analysed in detail by \citet{TARG2013pre} and was reviewed here and extended, focusing on 1D systems. The main features are the occurrence of localized states even when the branch of periodic states bifurcates supercritically from the uniform state, strongly subcritical bifurcations generating spatially localized states that may exist even outside the two-phase coexistence region between the uniform (liquid) state and the periodic (crystal) state, organization of these states into a slanted snakes-and-ladders structure, and a transition from slanted snaking to smooth snaking when the effective temperature is increased. In the latter case, the localized states increase smoothly in size without specific nucleation events. In fact, all these are properties of conserved systems {\it in general} \citep{Knob2016ijam}.
Moreover, stability still changes along the localized state branches owing to the continued existence of rung branches of asymmetric states. These properties are intrinsically related to the redistribution of the conserved density between patterned and uniform regions, i.e., it is a finite size effect. For an analysis of the transition from finite-domain bifurcation diagrams to the Maxwell construction of the thermodynamics of first order phase transitions in 1D and 2D see \citet{TFEK2019njp}. We have also explained how slanted snaking turns into aligned snaking when presenting the data over a different control parameter and have pointed out the resulting intricacies in the interpretation of the bifurcation diagram. These points are further elaborated in the conclusion of \citet{TARG2013pre}. Related connections between steady solutions of conserved and nonconserved variants of other models, namely, Allen-Cahn vs.\ Cahn-Hilliard model and Swift-Hohenberg vs.\ PFC model are discussed by \citet{EGUW2019springer}.

We have also considered the active PFC model by \citet{MeLo2013prl} where the concentration field is coupled to a polarization field in a manner that breaks the gradient dynamics form of the model, i.e., renders it ``active''. We have summarized and expanded on the main results of \citet{OpGT2018pre} for this model. In particular, we have illustrated how the slanted homoclinic snaking structure of the PFC case is amended by activity and showed that increasing activity beyond a threshold value leads to the onset of drift of the various localized states. Using the activity parameter as the main control parameter we explained that traveling localized states emerge either via drift-pitchfork bifurcations of steady symmetric localized states or via drift-transcritical bifurcations of steady asymmetric localized states. This is a particular feature of the active PFC model that has not yet, to our knowledge, been remarked upon. This onset behavior differs from findings for the nonvariational Swift-Hohenberg equation considered \citet{HoKn2011pre}. There all asymmetric states drift for any nonzero activity. 
The details of the present onset behavior, the various traveling states and their interactions are further discussed by \citet{OpGT2018pre} and~\citet{OKGT2020c}. The 2D case is considered by \citet{OKGT2020pre}.

Next, we have considered a two-field PFC model for a crystallizing binary mixture, e.g., of two species of colloids, first with a variational coupling as in \cite{HoAT2020jpcm} and then with a nonvariational coupling. In the variational case the system allows for various liquid and crystalline phases and we have explored the occurrence of localized states in the vicinity of first order phase transitions between the uniform (liquid) phase and a periodic (crystalline) alloy phase on the one hand, and a single-species crystalline phase and the crystalline alloy phase on the other. In both cases slanted snaking of localized states occurs and in fact represents a signature of the phase transition in the bifurcation diagrams for finite size domains. Here, we believe that a systematic study of bifurcation diagrams in the context of phase transitions may be useful for obtaining a deeper understanding of complex phase behavior in finite size systems.

The variational case serves as a reference for the two-field PFC model with a nonvariational coupling. The chosen coupling respects both conservation laws, i.e., it does not represent a chemical reaction between components, and can be seen as a continuous predator-prey-type model of two potentially crystallizing species: e.g., for positive nonvariational coupling $\alpha$ in Eqs.~\eqref{eq:cPFC} species 1 would be repelled by species 2 while species 2 is attracted by species 1, not unlike the attraction-repulsion set-up by \citet{BuFH2014sjads} and~\citet{ChKo2014jrsi}. Recently, such couplings have also been studied for active two-field Cahn-Hilliard-type models \citep{YoBM2020pnas,SaAG2020arxiv,FrWT2020arxiv}, for localized states in such models see \citet{FrTh2020arxiv} in the present volume.

Here, we have considered how activity affects homoclinic snaking in this type of two-field PFC model and have found that it acts much like the effective temperature in passive PFC models. However, depending on the sign of the active coupling, the ``destruction'' of the snaking structure with increasing activity changes. We have not explored the emergence of traveling and oscillating states but mention that such states occur in this system and promise rich bifurcation behavior that awaits exploration.

Finally, we have briefly analyzed another active PFC model for a binary mixture of a passive and an active colloid species as studied experimentally, see e.g. \citet{BDLR2016rmp}. Here we have coupled a passive single-species PFC model with an active single-species PFC model. Our analysis showed that depending on the type of underlying phase transition considered, snaking branches of localized states may or may not be destroyed by increasing the activity. Note also that in this model activity results in the emergence of traveling states that are not considered here.

Overall we have found that for all the considered PFC systems slanted snake-and-ladder structures of intertwined branches of symmetric localized states connected by branches of asymmetric localized states occur. In the variational case, i.e., for gradient dynamics, the homoclinic snaking structure is always an expression of a underlying first order phase transition and can be used to identify its equivalent for systems of finite size. In this case, by definition all states are at rest. For the one-field PFC model the transition from slanted homoclinic snaking to the thermodynamic Maxwell construction is considered by \citet{TFEK2019njp}. However, for the passive two-field PFC model where also three-phase coexistence occurs \citep{HoAT2020jpcm} this is more intricate and not yet done.

All the active models considered here are intrinsically out-of-equilibrium models as they feature nonreciprocal interactions \citep{IBHD2015prx} between the fields. As activity is increased from zero, first, the entire slanted snake-and-ladder branch structure of resting localized states remains qualitatively unchanged. In all the considered cases, increasing activity has an influence on the bifurcation structure similar to the effective temperature in the passive case. This was already remarked on by \citet{MeLo2013prl} when considering the occurrence of periodic crystalline structures in the one-field PFC model. For the symmetric localized states the existence as resting states is expected. However, for the asymmetric localized states their universal existence for all three considered active models is unexpected. It is explained in a forthcoming publication.

In the present work we have focused on the steady states and have mentioned traveling and oscillating localized states only in passing. As one increases the activity beyond threshold values such time-dependent states occur in all the considered active models. For the one-species active PFC model, the occurrence of drift bifurcations is studied by \citet{OpGT2018pre} and~\citet{OKGT2020pre}, and some oscillatory states are described by \citet{OKGT2020c}. However, the emergence and interplay of drift bifurcations and Hopf bifurcations on the snake-and-ladder structure still awaits more detailed studies.


\section*{Acknowledgement}

All authors thank the Center of Nonlinear Science (CeNoS) of the University of M{\"u}nster for support of their collaboration.
UT, LO and MH acknowledge support by the doctoral school ‘‘Active living fluids’’ funded by the German French University (Grant No. CDFA-01-14). The work of EK was supported in part by the National Science Foundation under grant DMS-1908891.



\end{document}